\documentclass[preprint2]{aastex}
\shorttitle{Variability and star formation in Leo~T}
\shortauthors{Clementini et al.}
\usepackage{url}

\begin{document}

\title{Variability and star formation in Leo~T,  the lowest luminosity star--forming galaxy known today\altaffilmark{1}}
\author{Gisella Clementini\altaffilmark{2}, Michele Cignoni\altaffilmark{2,3},  Rodrigo Contreras Ramos\altaffilmark{2,3}, Luciana Federici\altaffilmark{2}, 
Vincenzo Ripepi\altaffilmark{4}, Marcella Marconi\altaffilmark{4}, Monica Tosi\altaffilmark{2}, 
\and 
Ilaria Musella\altaffilmark{4}
}
\altaffiltext{2}{INAF, Osservatorio Astronomico di Bologna, Bologna, 
Italy; gisella.clementini@oabo.inaf.it, rodrigo.contreras@oabo.inaf.it, luciana.federici@oabo.inaf.it,monica.tosi@oabo.inaf.it}
\altaffiltext{3}{Dipartimento di Astronomia, Universit\`a di Bologna, Bologna, Italy; michele.cignoni@unibo.it}
\altaffiltext{4}{INAF, Osservatorio Astronomico di Capodimonte, Napoli, Italy; ripepi@na.astro.it, marcella@na.astro.it, ilaria@na.astro.it}
\altaffiltext{1}{Based on archive data collected with the Wide Field Planetary Camera 2 on board the
Hubble Space Telescope.}

\begin{abstract}
We present results from the first combined study of variable stars and
star formation history (SFH) of the Milky Way (MW) ``ultra-faint" dwarf
(UFD) galaxy Leo\,T, based on F606W and F814W multi-epoch archive
observations obtained with the Wide Field Planetary Camera 2 on board
the Hubble Space Telescope.  We have detected 14 variable stars in the
galaxy. They include one fundamental-mode RR Lyrae star and 10 Anomalous
Cepheids with periods shorter than 1 day, thus suggesting the
occurrence of multiple star formation episodes in this UFD, of which
one about 10 Gyr ago produced the RR Lyrae star.  A new estimate of
the distance to Leo~T of 409 $^{+29}_{-27}$ kpc (distance modulus of
23.06 $\pm$ 0.15 mag) was derived from the galaxy's RR Lyrae star.  Our
$V, V-I$ color-magnitude diagram of Leo\,T reaches $V \sim$ 29 mag and
shows features typical of a galaxy in transition between dwarf
irregular and dwarf spheroidal types. A quantitative analysis of the
star formation history, based on the comparison of the observed $V,
V-I$ CMD with the expected distribution of stars for different
evolutionary scenarios, confirms that Leo\,T has a complex star
formation history dominated by two enhanced periods about 1.5 and 9 Gyr
ago, respectively.  The distribution of stars and gas shows that the
galaxy has a fairly  asymmetric structure. 
\end{abstract}


\section{Introduction}
Numerical simulations and semi-analytical models operating in the
$\Lambda$-cold-dark-matter ($\Lambda$-CDM) scenario of galaxy
formation (e.g. Bullock \& Johnston 2005) suggest that the halos of
large spirals like the Milky Way (MW) and the Andromeda (M31) galaxies
could have entirely been built up by merging of disrupted satellites.
Despite an intensive search, the surviving ``building blocks" of this
formation process so far have remained elusive.  However, new hope to
the quest has been triggered in the last few years, by the discovery
of many faint dwarf galaxies in the outskirts and halos of the MW and
M31 spirals.  The census of the new MW companions counts so far 17
newly discovered dwarf galaxies, that were detected mainly from the
analysis of the Sloan Digital Sky Survey (SDSS) data (see e.g.
Belokurov et al. 2007, 2010; Watkins et al. 2009; Koposov et al. 2009,
and references therein).  Similarly, 12 dwarf galaxies were known to
be M31 companions until 2004, of which only 6 are dSphs, but a wealth
of 19 previously undetected new satellites were discovered in the last
5 - 6 years, by the panoramic surveys of the M31 halo carried out with
the Isaac Newton and the Canada-France-Hawaii Telescopes (Ferguson et
al. 2002; Ibata et al. 2007; McConnachie et al. 2009; Richardson et
al. 2011) and, more recently, also by the SDSS (Slater et al. 2011; 
Bell et al. 2011).

\begin{figure*}
\begin{center}
\includegraphics[width=12cm,bb=45 185 535 665,clip]{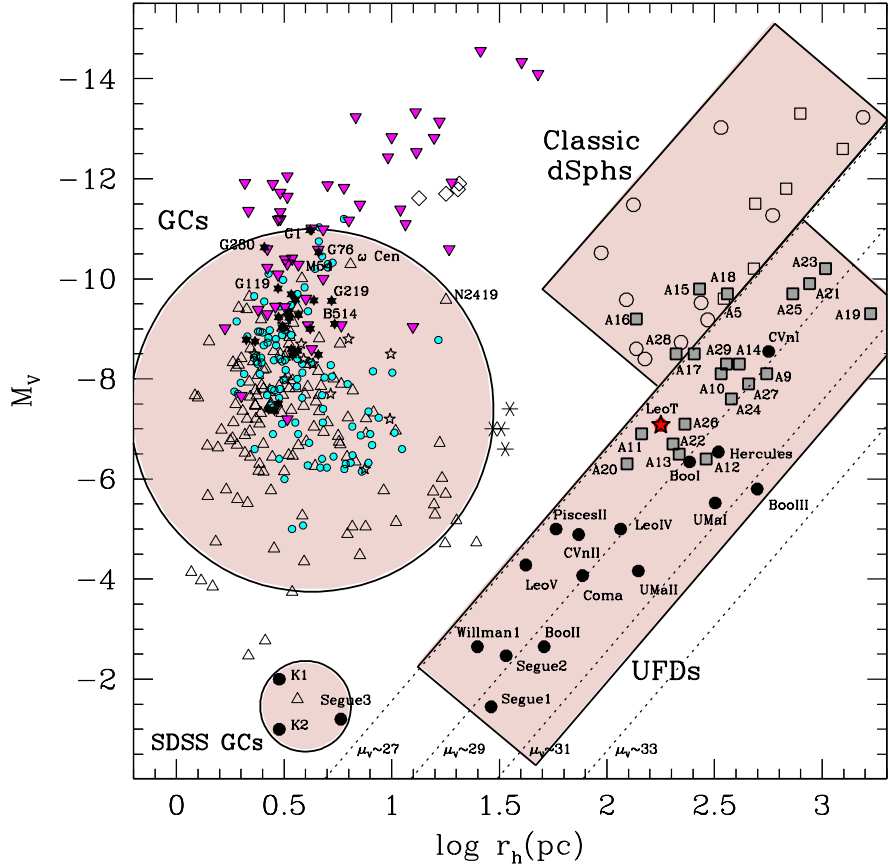}
\caption{Absolute magnitude (${\rm M_V}$) {\it versus} half-light
  radius ($r_h$) for different objects. Lines of constant surface
  brightness are marked. Open circles and squares are the classical
  dSphs surrounding the MW and M31, respectively.  Filled circles are
  the MW UFDs, and 3 extremely low luminosity GCs, discovered mainly
  from the analysis of the SDSS data (see e.g. Belokurov et al. 2007,
  2010; Watkins et al. 2009; Koposov et al. 2009; and references
  therein). The Leo~T UFD is marked by a large filled star (red, in the electronic edition of 
  the journal).  Filled squares (grey) are the M31 dSph satellites discovered after 2004
  (see, e.g., Richardson et al. 2011, and references therein; Slater
  et al. 2011; Bell et al. 2011). Open triangles are the Galactic GCs
  from Harris (1996), and Mackey et al. (2006).  Small filled stars are the
  GCs of the Andromeda galaxy, from Federici et al. (2007).  Small open
  stars are the GCs in the outer halo of M31, from Mackey et
  al. (2007) and Martin et al. (2006). Asterisks are the extended M31
  GCs, from Huxor et al. (2005) with parameters measured by Mackey et
  al (2006).  Filled circles (cyan) are the GCs of NGC~5128, from
  McLaughlin et al. (2008). Filled  inverted triangles (magenta) are the
  nuclei of dwarf elliptical galaxies in the Virgo cluster, from Cot\'e
  et al. (2006). Diamonds are the ultra-compact dwarfs in the Fornax
  cluster, from De Propris et al. (2005).}
\label{figbelokurov}
\end{center}
\end{figure*}

The new dSphs are fainter than those previously known, with
typical surface brightness generally around 28 mag/arcsec$^2$ or less, 
hence they were named ``ultra-faint" dwarfs (UFDs).  With
luminosities that reach as low as 10$^3$ L$_{\odot}$, the UFDs provide
the ultimate opportunity to test models of the formation and chemical
enrichment of the first bound structures, and their implications for
the formation of larger galaxies (e.g. Bovill \& Ricotti 2011a,b and references therein;  
Tumlinson 2010).  Fig. 1 shows the location of
classical (bright) dSphs and UFDs in the absolute magnitude versus
half-light radius (${{\rm M_V} - \log r_h}$) plane.  The plot is an
adapted and updated version of Belokurov et al.'s (2007) Figure 8.
The MW and some of the M31 GCs are also shown in the plot, for
comparison, as well as the GCs of NGC~5128 and the ultra compact dwarf
ellipticals in the Virgo cluster.  Likely due to an observational
selection effect, the Andromeda's new satellites are generally
brighter than the MW UFDs. Nevertheless, with faint luminosities
typical of the bulk of globular clusters (GCs) and large spatial
dimensions typical of dSphs, both the MW and the M31 new
satellites sample a totally unexplored region of the ${{\rm M_V} -
  \log r_h}$ plane (see Fig.~\ref{figbelokurov}).

The UFDs appear to be clustered in groups on the sky (see, e.g.,
Figure 1 of Richardson et al. 2011).  Their velocity dispersions are
small, 3-4 km/s, and they are quite extended (see
Fig.~\ref{figbelokurov}). The MW UFDs seem to be particularly dark
matter dominated (see Table~1 of Wolf et al. 2010, and references
therein) and their chemical abundances are extreme, with large
dispersions, and stars as metal poor as [Fe/H]=$-4$ (see Tolstoy, Hill
\& Tosi, 2009 and references therein).  Often they have irregular
shape or are elongated, as Canes Venatici II, Segue I, Ursa Major II,
Hercules (see Fig.5 of Belokurov et al. 2007; Fig. 11 of Mu{\~n}oz et
al. 2010; and Figs. 6 and 7 of Musella et al. 2012), as likely
distorted by the tidal interaction with the MW.  Some of them (Bootes
II and Segue I) seem to be entangled in complex kinematic streams as
the Sagittarius tails (see, map on V. Belokurov's web page:
\url{http://www.ast.cam.ac.uk/~vasily/sdss/field_of_streams/dr6/fos_dr6_marked_names.tif}),
see Law \& Majewski (2010) for an in depth discussion of this issue.
All UFDs host an ancient population as old as about 10 Gyr, and
generally have GC-like CMDs, resembling, although more dispersed, the
CMDs of metal poor Galactic clusters like M92, M15 and M68 (Belokurov
et al. 2007; Moretti et al. 2009; Musella et al. 2012; Brown et
al. 2012).

The only remarkable exception to this general behavior is represented
by the Leo~T UFD (Irwin et al. 2007).  Discovered as a stellar
overdensity in the SDSS Data Release 5, the galaxy's CMD, based on
follow-up observations with the 2.5 m Isaac Newton Telescope,  
revealed that Leo~T is characterized by an intermediate-age
stellar population with a metallicity of [Fe/H] $\sim -1.6$ dex,
together with a young population of blue stars of age $\sim$ 200 Myr
(Irwin et al. 2007). These authors estimated for the galaxy a distance
modulus of $(m-M)_0 = 23.1$ mag (corresponding to a distance D= 417
kpc), an half-light radius $ r_h$ (Plummer\footnote{Irwin et al.'s (2007) half-light radius is 
derived by fitting the radial profile of Leo~T with a standard Plummer law (Plummer 1911).})=1.4$^\prime$, a surface
brightness $\mu_{0, V}$ (Plummer)= 26.9 mag/arcsec$^2$, and an integrated
magnitude $M_{tot, V} = -$7.1 mag.  According to these parameters the
galaxy locates at the bright end of the MW UFD distribution in the
${{\rm M_V} - \log r_h}$ plane (see Fig.~\ref{figbelokurov}), ranking
second only to Canes Venatici~I (CVn~I), and in the transition region
between the MW and the M31 UFDs.  The deeper ($g \lesssim$ 26.5 mag)
CMD published by de Jong et al. (2008) confirms the presence in Leo~T
of very young stars with ages between $\sim$ 200 Myr and 1 Gyr, as
well as an older stellar population ($>$5 Gyr and [Fe/H]$\sim -1.7$
dex).  The galaxy is embedded into an HI cloud with an heliocentric
velocity of 38.6 km s$^{-1}$, a velocity dispersion of 6.9 km s$^{-1}$
and an estimated mass of 2.8 $\times 10^5 M_{\odot}$ (Ryan-Weber
et al. 2008), or 4.3$\times 10^5 M_{\odot}$ (Grcevich \& Putman 2009).
Leo~T remains, so far, the only UFD found to contain a significant
amount of neutral gas. The presence of HI above a certain column
density is often associated with star formation, and indeed Leo T is
the only UFD, and the lowest luminosity galaxy, with ongoing star
formation known to date. The presence of gas also suggests that Leo~T
may be affected by internal differential reddening.

Simon \& Geha (2007) obtained Keck DEIMOS spectra of 19 red giants in
Leo~T for which they estimated a mean metal abundance of [Fe/H]=$-2.29
\pm 0.10$ (on the Rutledge et al. 1997 metallicity scale) with a
dispersion of $\sigma_{\rm [Fe/H]}$=0.35 dex, from the equivalent
width of the Ca triplet absorption lines. Kirby et al. (2008) have
re-analyzed Simon \& Geha (2007) spectroscopic material for Leo~T by
applying an automated spectral synthesis technique.  They derived
metallicities in the range of [Fe/H]=$-$0.12 to $-3.22$ dex, with an
average value of $\langle {\rm [Fe/H]} \rangle $ = $-2.02 \pm 0.05$
dex and dispersion $\sigma_{\rm [Fe/H]}$=0.54 dex (where individual
stellar [Fe/H] values were first weighted by the inverse square of the
errors and then averaged to obtain the mean [Fe/H] value).  This value
was later revised to $\langle {\rm [Fe/H]} \rangle $ = $-1.99 \pm
0.05$ dex, $\sigma_{\rm [Fe/H]}$=0.52 dex, by Kirby et al. (2011).
The spectroscopic mean metallicities are lower than the values adopted
by Irwin et al. (2007) and de Jong et al. (2008).  Simon \& Geha
(2007) also measured a mean stellar velocity of 38.1 $\pm$ 2.0 km
s$^{-1}$ with a dispersion of 7.5 $\pm$ 1.6 km s$^{-1}$, in excellent
agreement with the HI velocity and gas velocity dispersion of 6.9 km
s$^{-1}$ measured by Ryan-Weber et al. (2008). From the velocity
dispersion Simon \& Geha (2007) infer a dark halo mass of $(8.2 \pm
3.6) \times 10^6 M_{\odot}$ and a mass-to-light ratio of $138 \pm 71
M_{\odot}/L_{\odot,V}$ for Leo T, while from the HI observations,
Ryan-Weber et al. (2008) infer a lower limit for the total dynamical
mass of $\sim 3 \times 10^6 M_{\odot}$ and a mass-to-light ratio of
$\gtrsim 56 M_{\odot}/L_{\odot,V}$. 

Rocha et al. (2012) have estimated infall times ($t_{infall}$) for the
MW satellites based on the Via Lactea II cosmological simulation
(Diemand, Kuhlen \& Madau 2007; Diemand et al. 2008; Kuhlen
2010). They find that Leo T was accreted much more recently
($t_{infall} \leq$ 1 Gyr) than the other MW dSph and UFD satellites
(see Figure 3 of Rocha et al.'s paper) and would be just falling into
the Milky Way for the first time.  This could explain why Leo~T
managed to retain its gas and kept forming stars until  
a few hundreds Myr ago.

As part of a project aimed at understanding the evolution of the UFDs,
their connection with the MW and M31, and with the classical dwarfs
and the GCs, we have already studied 7 MW UFDs (Bootes~I,
Dall'Ora et al. 2006; CVn~I, Kuehn et al. 2008;
Canes Venatici~II, Greco et al. 2008; Coma, Musella et al. 2009;
Leo~IV, Moretti et al. 2009; Ursa Major~II, Dall'Ora et al. 2012; and
Hercules, Musella et al. 2012), and in this paper we
present results from the combined study of variable stars, star
formation history (SFH), and spatial distribution of the resolved
stellar populations in Leo~T.
 \begin{figure*}
\includegraphics[width=12cm,clip]{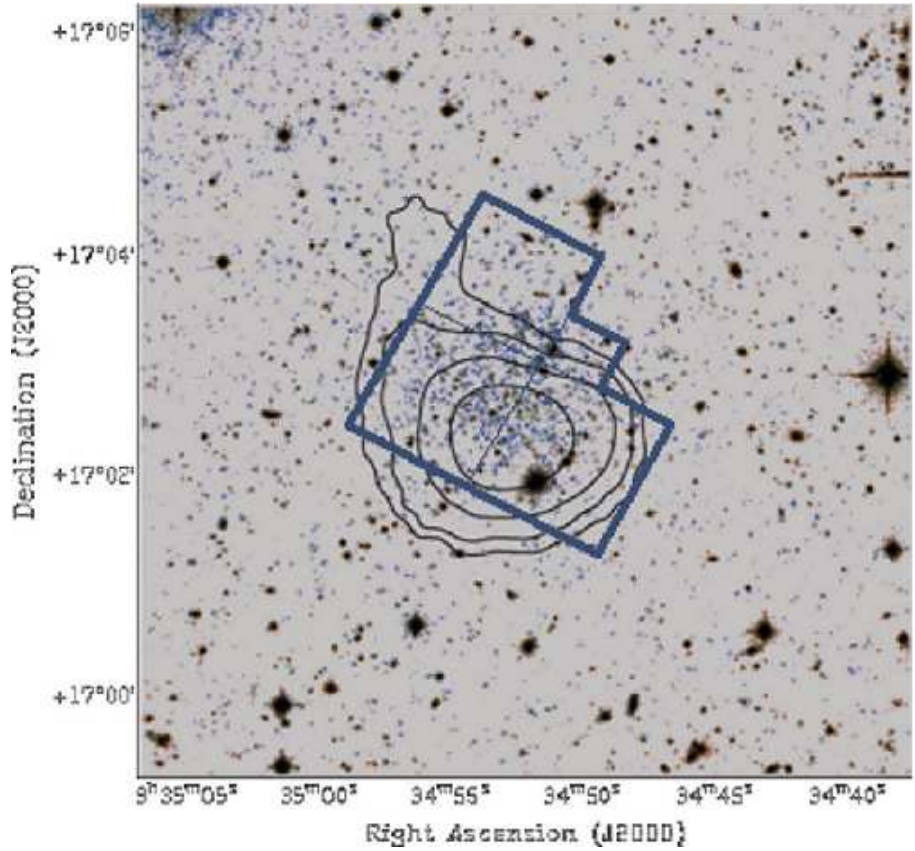}
\caption{Map of the Leo~T UFD showing the gas distribution, according
  to Ryan-Weber et al. (2008) and the position of the field observed
  with the HST. Adapted from Ryan-Weber et al. (2008) Fig. 1.}
\label{figHImap}
\end{figure*}
The properties of the Leo~T variable stars provide us information on
the conditions at the epochs of formation of these variables (young
ages: 50 -- 200 Myr, from short and intermediate-period Classical
Cepheids $-$ CCs; intermediate-age: $\sim$ 1 -- 2 Gyr, from 
the Anomalous Cepheids $-$ ACs; and old ages: t $>$ 10 Gyr, from the
RR Lyrae stars) and, combined with deep CMDs of the stars resolved on
the stacked images, have allowed us to perform a quantitative analysis
of the SFH history occurred in Leo~T, using the synthetic CMD method
(Cignoni \& Tosi 2010, and references therein).

Observations, data reduction and calibration of the Leo~T photometry
are presented in Section 2. Results on the variable stars, the
catalogue of light curves, and the distance to the galaxy they lead to 
are discussed in Section 3.  The CMD of Leo\,T
and the galaxy SFH are presented in Sections 4 and 5, respectively.
The comparison of the spatial distribution of the Leo~T gaseous and
stellar components is discussed in Section 6.  Finally, a summary of
the main results is presented in Section 7.
\section{Observations and Data Reduction}
\begin{table*}
 \begin{center}
      \caption[]{Log of the observations}
	 \label{tab1}
	 $$
	\begin{array}{lccccc}
	   \hline
	    \hline
	   \noalign{\smallskip}
{\rm ~~~~~Dates}& {\rm JD~interval}&{\rm Filter}&{\rm Exposure~lenght}&{\rm N}&{\rm Total~ exposure}\\
     	        & (-2454000)                        &                  &                                         &            & \\
     	        &  { \rm (d)}              &            & {\rm (sec)}         &  & {\rm (sec)} \\
 	    \noalign{\smallskip}
	    \hline
	{\rm Oct. 21, Dec.6, 2007}& 395.17-441.08 &  {\rm F606W}& 1200 & 16& 19200 \\
        {\rm Oct. 29-30, Dec.8-9, 2007}  & 403.36-444.08 &  {\rm F814W}& 1200 & 26& 31200 \\
\hline
	 \end{array}
	 $$
	 \end{center}
	 \normalsize
         \end{table*} 

The multi-epoch images of Leo~T [RA(J2000)= 09$^h$ 34$^m$ 53.4$^s$;
  DEC(J2000)=+17$^{\circ}$~ 03$^{\prime}$~05$^{\prime \prime}$, $l$
  =214.9$^{\circ}$, $b$=43.7$^{\circ}$, Irwin et al. 2007] used in
this paper were retrieved from the HST archive.  They were obtained
with the Wide Field Planetary Camera 2 (WFPC2) from October 21 to
December 9, 2007 as part of GO program 11084 (PI.: D. Zucker), and
consist of 16 F606W (V) and 26 F814W (I) frames, each corresponding to
a 1200 sec exposure, thus making total exposure times of 19,200 sec and
31,200 sec, respectively.  Specifically, the sixteen observations in
F606W were obtained on 2007 October 21 and December 6, in two blocks
of 4-orbits each, with a $\sim$ 52 min gap occurring between consecutive
orbits in each block and each orbit filled by a pair of 1200 sec 
exposures in the same filter.  The twenty-six F814W observations were
obtained in 3 blocks, on 2007 October 29-30, December 8 and December
9. The first one comprising 5 orbits, while the other two 4 orbits
each. As in the previous case, two exposures of 1200 sec  in the same
filter were acquired in each orbit, and a gap of $\sim$ 52 minutes
was used to separate consecutive orbits.  A log of the observations is
provided in Table~\ref{tab1}.  Fig.~\ref{figHImap} shows the position
of the four cameras of the WFPC2 on an image of Leo~T published by
Ryan-Weber et al. (2008) with overlaid the HI contours. The HST
observations were centered on Irwin et al. coordinates for the
galaxy. This roughly corresponds to the common vertex of the 4 WFPC2
cameras. The $2.6^{\prime} \times 2.6^{\prime}$ field of view (FOV) of
the WFPC2 allows to cover almost entirely the region within the
half-light radius of Leo~T.

\begin{figure*}
\begin{center}
\includegraphics[width=12cm,clip]{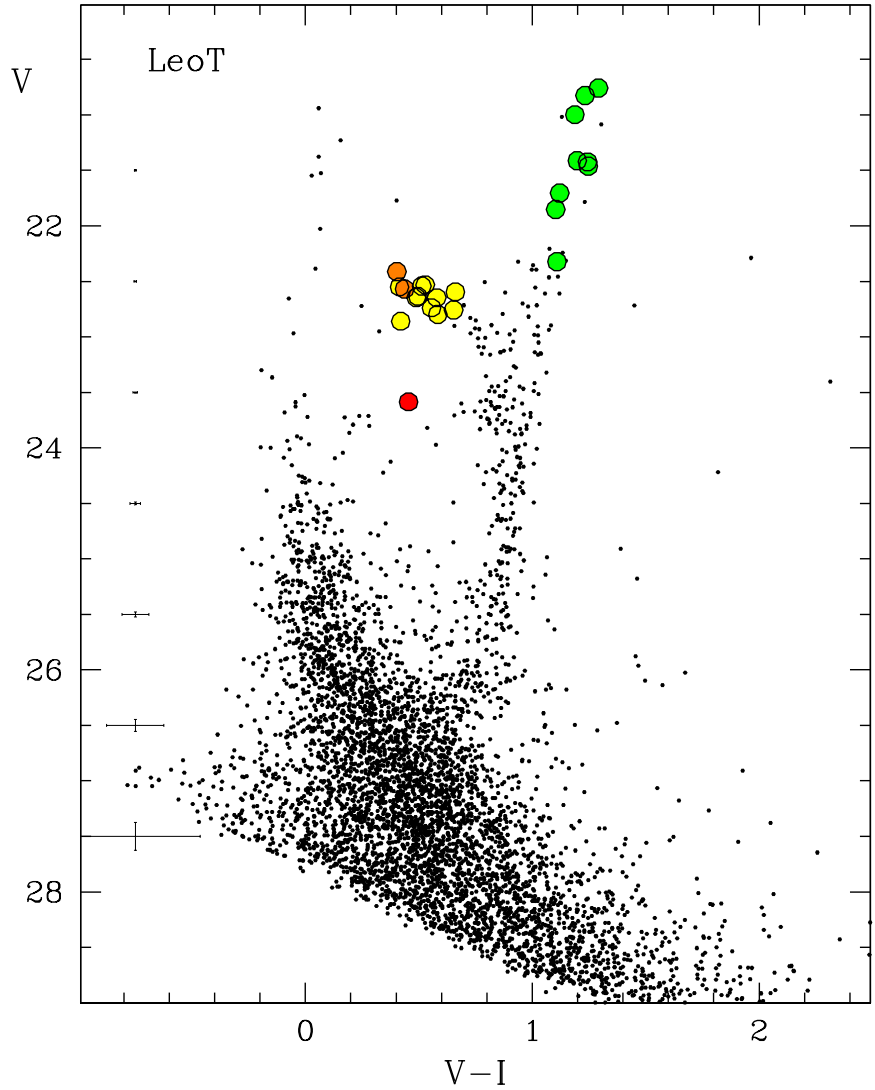}
\caption{$V, V-I$ CMD of Leo T from the HST WFPC2 data.
The  RR Lyrae star is marked by a filled circle (in red, in the electronic version of the paper), the variable stars brighter that the HB by (yellow) filled circles. 
Among them, the (orange) filled circles are suspected binaries. The variables are plotted in the CMD according to their intensity-averaged 
magnitudes and colors. (Green) filled circles are Leo~T red giant stars studied  spectroscopically by Kirby et al. (2008).}
\label{figCMD1}
\end{center}
\end{figure*}

The photometric reductions of the individual pre-reduced images
supplied by the STScI pipeline were done using HSTphot (Dolphin et
al. 2000). HSTphot performs point-spread function (PSF) photometry
using PSFs especially tailored to the HST/WFPC2 camera and accounts
for both hot-pixels and cosmic-rays using the information provided in
each data quality image, and the comparison of multiple exposures, respectively. 
 HSTphot also fits the sky locally around
each detected source, and provides magnitudes of the measured sources
corrected for charge-transfer efficiency (CTE, Dolphin 2009), and
transformed to the HST flight system and the Johnson-Cousins photometric system
using the transformations available at ${\rm http://purcell.as.arizona.edu/wfpc2\_calib/}$. 
We used the
quality information provided by the HSTphot package to clean the list of
detected sources, selecting for the CMD only stellar detections
with valid photometry (object type flag=1) on all input images,
global sharpness parameter $-0.3 \leq sharpness \leq $0.3, and $\chi^2
\leq$ 1.1, in each filter.  The $V, V-I$ CMD of Leo~T obtained with
this procedure is shown in Fig.~\ref{figCMD1}.
Typical internal errors of the combined photometry for non-variable
stars at the magnitude level of the Leo~T horizontal branch (HB, $V
\sim 23.6$ mag) are $\sigma_{V}$ = 0.005 mag and  $\sigma_{I}$= 0.006 mag,
respectively.

\section{Variable Stars}
The identification of the variable stars was carried out using the
F606W and F814W time-series data separately, and a number of different
approaches. First, candidates were identified from the scatter
diagrams of the F606W and F814W datasets, using VARFIND, a custom
software developed at the Bologna Observatory by P. Montegriffo.  The
second approach was to use the Stetson variability index (Stetson 1994).  Then, as a third approach and to detect potential
variables fainter than the galaxy horizontal branch (HB), an
independent search was performed on the time-series data of all
sources brighter than $V$=25.5 mag ($\sim$ 2500 objects in total),
that were analysed in both filters by means of a Lomb-Scargle
algorithm (Press \& Rybicki 1989) to find the most probable periods.
The light curves of the candidate variables found with the above
procedures were then inspected visually.  Particular care was devoted
to check stars brighter than $V$=22.5 mag and with color $V-I \leq$
0.5 mag, that, according to the comparison with the isochrones (see
Section 4) might be crossing the instability strip during the blue
loop phase of their evolution.  At the end of the whole search process
we were left with a list of about a hundred candidates, of which 14
were eventually confirmed.  Final periods for these confirmed variable
stars were obtained with the Grafical Analyser of Time Series
(GRATIS), a private software developed at the Bologna Observatory by
P. Montegriffo (see, e.g., Clementini et al. 2000), which uses both the
Lomb periodogram (Lomb 1976) and the best fit of the data with a
truncated Fourier series (Barning 1963). The final periods adopted to fold the light curves were those that minimize 
the r.m.s. scatter of the truncated Fourier
series best fitting the data.  
The scheduling of the Leo~T observations is not optimized for
variability studies, nevertheless, we were able to obtain reliable
light curves (particularly in the $I$ band), and periods for the
variable stars accurate to the third/forth decimal place, by iterating the period search procedure between the
F606W and F814W data. 
The variable stars are plotted on the
galaxy CMD in Fig.~\ref{figCMD1} using their intensity-averaged
magnitudes and colors, and different symbols for the different types
of variability.  We have also marked with (green) filled circles the RGB
stars measured spectroscopically by Kirby et al. (2008) that fall in
the WFPC2 FOV.  The light curves of the variable stars we have
identified in Leo~T are presented in Fig.~\ref{lc1}.
\begin{figure*} 
\begin{center}
\includegraphics[width=12cm,clip]{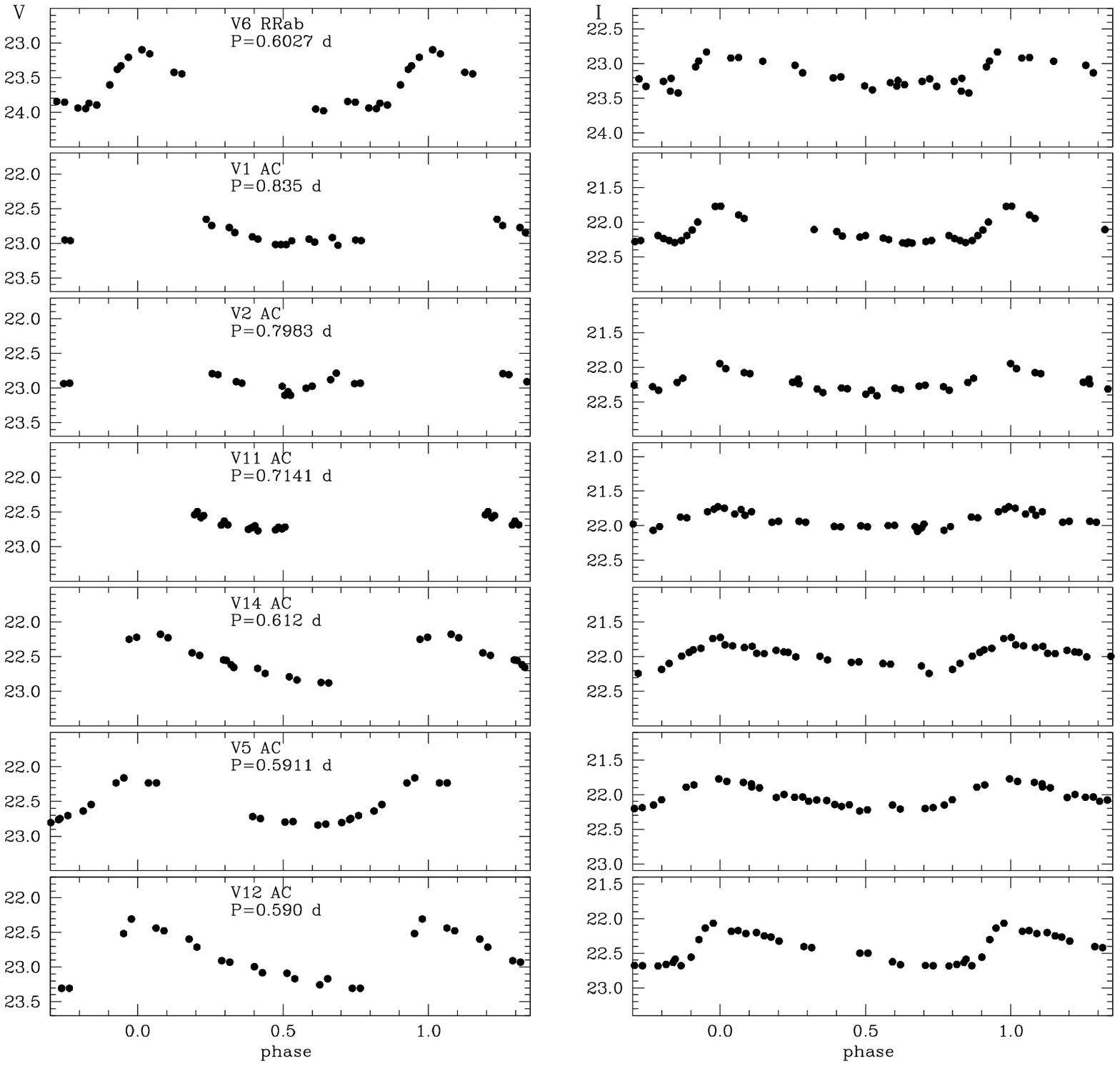}
\caption{$V$ (left panels) and $I$ (right panels) light curves of the
  RR Lyrae star (top panels) and ACs identified in Leo T.  Typical
  errors of the single data points for the RR Lyrae star are of about
  0.02 and 0.04 mag in $V$ and $I$, respectively; for the brighter
  variables they are of 0.01 mag for $V$ in the range of 22.8 to 22.5 mag,
  and between 0.03 and 0.02 mag for $I$  in the range of 22.4 to
  22.0 mag.}
\label{lc1}
\end{center}
\end{figure*}

\begin{figure*}
\begin{center}
\figurenum{4}
\includegraphics[width=12cm,clip]{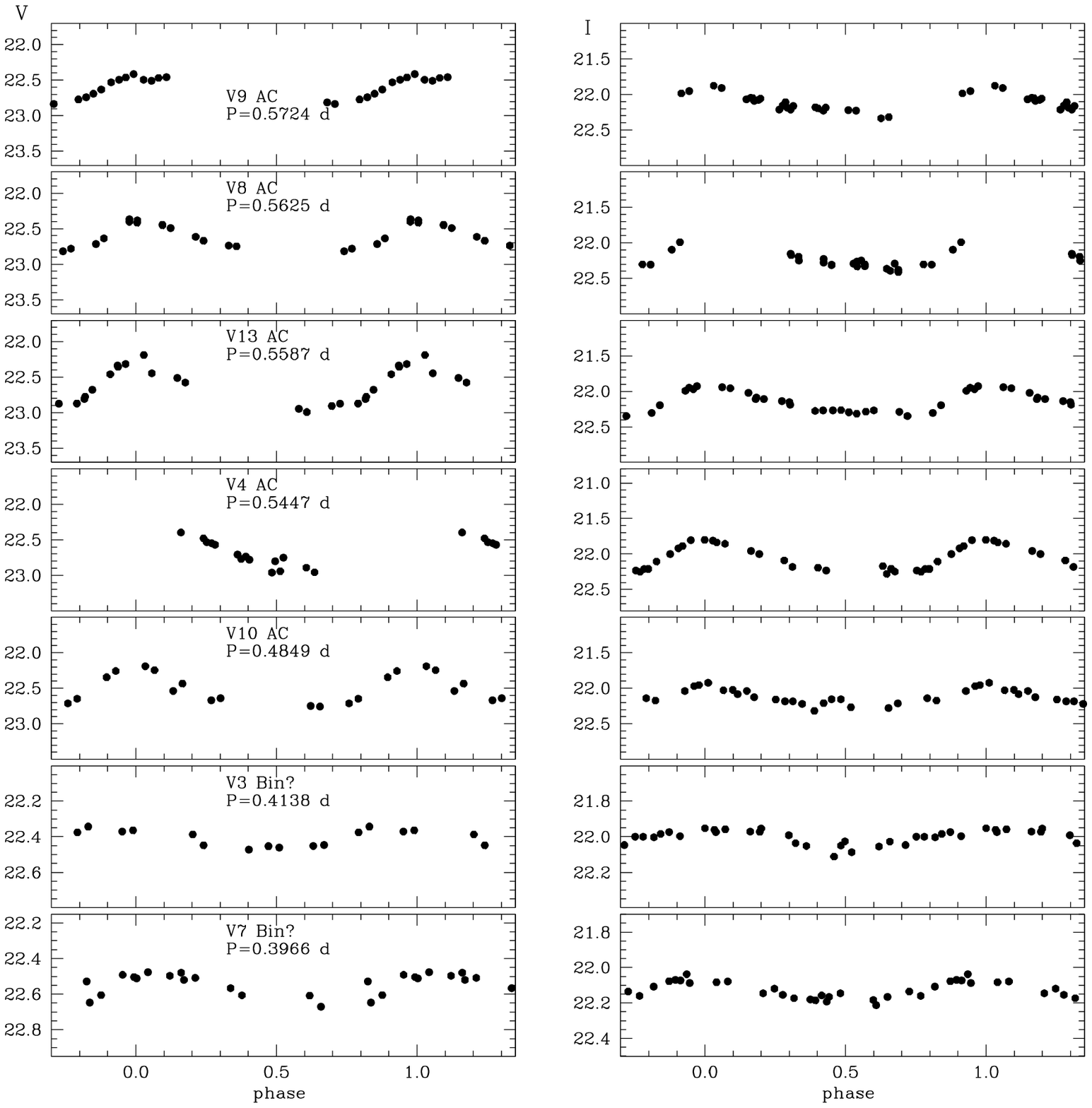}
\caption{ continued -- the upper five pairs of panels show  ACs, the lower two are suspected 
  binaries.  Note that, due to the small amplitude of their light variation  for the binaries we have used a Y-scale significantly smaller than for the ACs.}
\label{lc2}
\end{center}
\end{figure*}

\begin{figure*}
\includegraphics[width=14.0cm,clip]{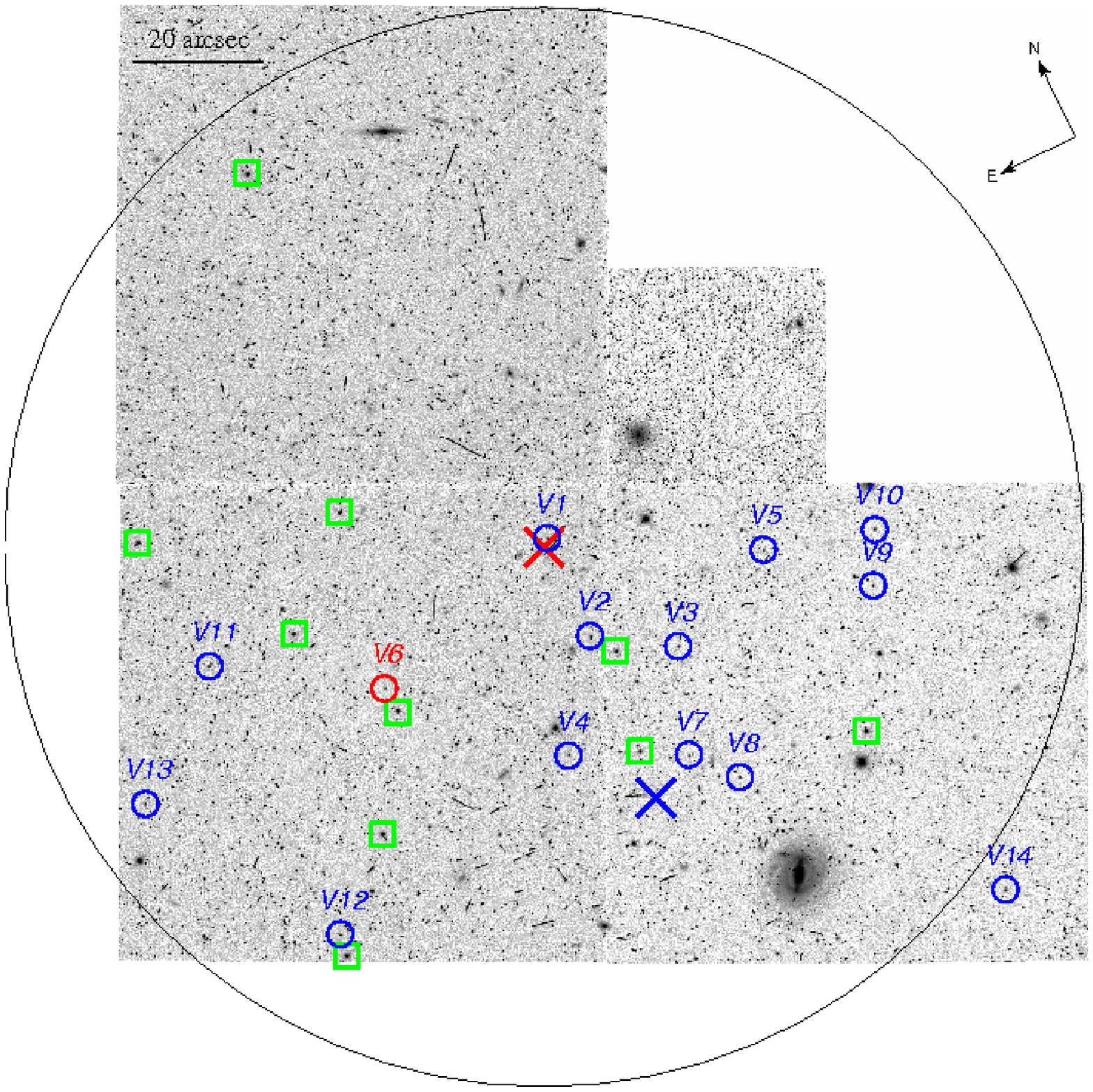} 
\caption{Image of Leo T on the 4 cameras of the WFPC2. The (red) cross 
  almost coinciding with the position of variable star V1, marks the
  center of Leo~T according to Irwin et al. (2007), and the large open
  circle is drawn with the galaxy half-light radius, 
  according to the same authors.  The RR Lyrae star  (V6) is marked by a (red)
  open circle, the (blue) open circles indicate  the variable stars brighter than the HB, among them V3 and V7
  are suspected binaries. (Green) open squares are the RGB stars
  measured spectroscopically by Kirby et al. (2008).  A (blue) cross marks the
  center of the HI distribution according to Ryan-Weber et
  al. (2008).}
\label{figmapHST4}
\end{figure*}

\begin{figure*}
\begin{center}
\includegraphics[width=14.0cm,clip]{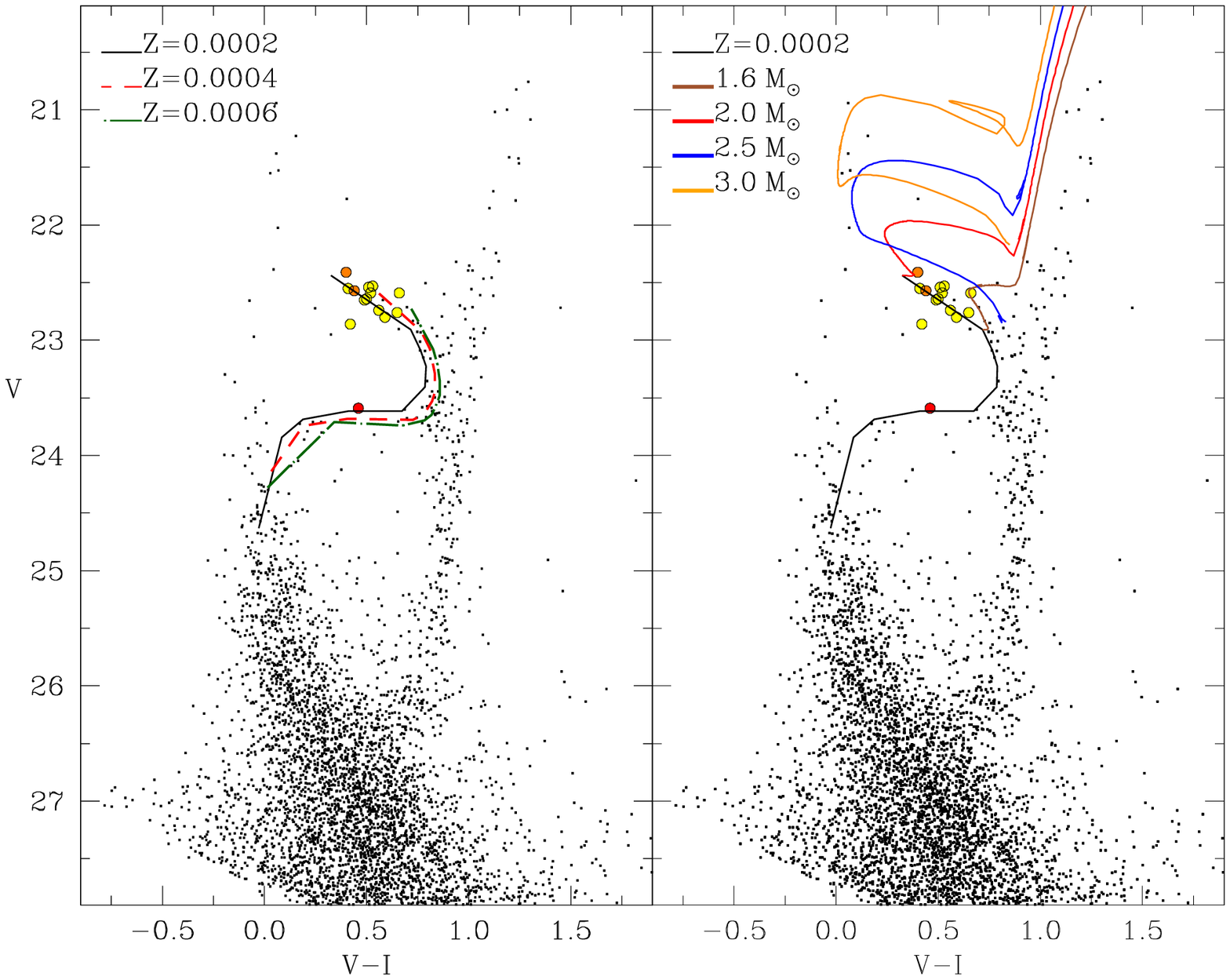}
\caption{{\it Left panel:} $V, V-I$ CMD of Leo~T with superimposed the
  ZAHBs of evolutionary models for metal abundances $Z$=0.0002 (black
  solid line), 0.0004 (red dashed line), and 0.0006 (green dot-dashed
  line), respectively, and masses in the range of 0.65 to 2.0
  M$_{\odot}$ from Cariulo et al. (2004, see text for details). The
  ZAHBs are plotted on the observed CMD assuming $E(B-V)$=0.03 mag
  (from Schlegel et al., 1998) and a distance modulus of 23.1 mag,
  according to Irwin et al. (2007).  {\it Right panel:} He burning and
  AGB evolution for stellar tracks between 1.6 and 3.0 $M_{\odot}$ and
  metallicity $Z$=0.0002. The $Z$=0.0002 ZAHB is also shown.}
\label{figCMD2}
\end{center}
\end{figure*}


\begin{figure*}
\begin{center}
\includegraphics[width=12.0cm,clip]{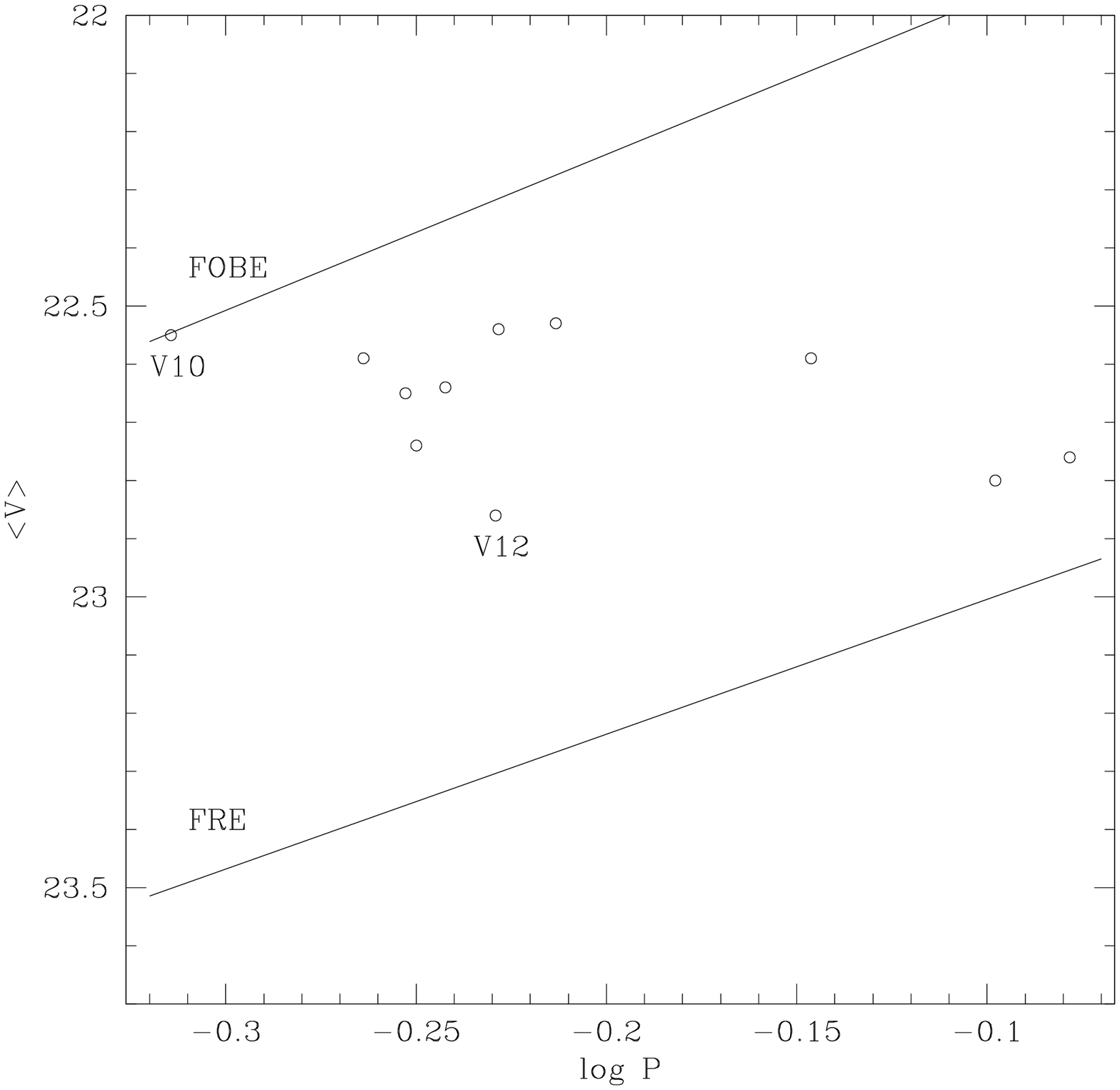}
\caption{Comparison of the blue (FOBE) and red (FRE) boundaries of the instability strip for ACs
  predicted by the nonlinear convective pulsation models of Marconi et
  al. (2004) with the distribution of the candidate AC stars observed
  in Leo~T.}
\label{AC_strip}
\end{center}
\end{figure*}

The Leo~T variables include one fundamental-mode RR Lyrae star with
period P=0.6027 day, and 13 variables about one magnitude brighter
than the HB with periods shorter than 1 day.  No candidate
variables fainter than the galaxy HB could be reliably confirmed, nor
was any variable star detected among the few blue loop objects in the
galaxy's CMD, hence no Classical Cepheids seem to be present in Leo~T,
indicating a negligible  star formation (SF) rate in the last half Gyr.
\begin{table*}
\scriptsize
\caption[]{Identification and properties of the Leo~T variable stars}
\label{tab2}
\begin{tabular}{lclclllllll}
\hline
\hline
\noalign{\smallskip}
{\rm Name}&{\rm Id}&~~~~~~~~{\rm $\alpha$}&{\rm $\delta$}&{\rm Type}&~~~~P&{\rm Epoch (max)}&~~$\langle V\rangle$&~~$\langle I\rangle$& A$_V$~~     
&A$_I$~~\\
~~    &       &~~~~~{\rm (2000)}&{\rm (2000)}&          & ~{\rm (days)}& ~($-$2454000)   & {\rm (mag)}            & {\rm (mag)}  &{\rm (mag)}&{\rm (mag)}\\ 
                    &{\rm (a)}&~~~~~~~{\rm (b)}       &{\rm (b)}      &          &                         &                               &          &   &    &                    \\
\noalign{\smallskip}
\hline
\noalign{\smallskip}
    {\rm V1 }  &7140 & 09:34:53.32  & +17:03:06.2 &~{\rm AC}  &0.835 & ~~444.010 & 22.76:  & 22.11  &  0.85:    & 0.54           \\ 
    {\rm V2 }  &6441 & 09:34:53.37  & +17:02:49.8 &~{\rm AC}  &0.7983& ~~442.8650& 22.80:& 22.21 &  0.63:    &0.36 \\ 
    {\rm V3 } &7112  & 09:34:52.56  & +17:02:42.1 &~{\rm Bin?}  &0.4138 & ~~403.430 & 22.41 & 22.01&  0.11    &0.11      \\ 
    {\rm V4 }  &6438 & 09:34:54.15  & +17:02:34.7 &~{\rm AC}  &0.5447& ~~444.0637& 22.59: & 22.07 & 0.78:   &0.45 \\ 
   {\rm V5 }  &7093  & 09:34:51.27  & +17:02:49.9 &~{\rm AC}  &0.5911& ~~443.999  & 22.54  & 22.03 &  0.69   &  0.44       \\ 
    {\rm V6 }  &\nodata & 09:34:55.63(c) & +17:02:56.5 &~{\rm RRab}&0.6027& ~~402.5277& 23.59 & 23.13 & 0.88 &    0.53 \\ 
    {\rm V7 }  &7118 & 09:34:52.99  & +17:02:26.3 &~{\rm Bin?}  &0.3966& ~~395.323 & 22.57  & 22.13 & 0.16  &  0.12  \\ 
    {\rm V8 }  &7108 & 09:34:52.60  & +17:02:19.8 &~{\rm AC}  &0.5625& ~~440.880 & 22.74: & 22.18:  &  0.55  & 0.35:  \\ 
    {\rm V9 }  &\nodata & 09:34:50.38(c) & +17:02:37.4 &~{\rm AC}  &0.5724& ~~443.912 & 22.64: & 22.14 & 0.61: &  0.47\\ 
    {\rm V10}  &7075 & 09:34:50.09  & +17:02:45.1 &~{\rm AC}  &0.4849& ~~442.875 & 22.55  & 22.14   &  0.60 &  0.29     \\ 
    {\rm V11 } &7212 & 09:34:57.23  & +17:03:11.8 &~{\rm AC}  &0.7141& ~~395.025 & 22.59: & 21.93 &  0.50 &  0.29      \\ 
    {\rm V12 } &6542 & 09:34:57.25  & +17:02:25.1 &~{\rm AC}  &0.590 & ~~395.20  & 22.86  & 22.44 &  1.03 &  0.66      \\ 
    {\rm V13 } &7235 & 09:34:58.52  & +17:02:56.8 &~{\rm AC}  &0.5587& ~~441.102& 22.65: & 22.16 &  0.74 &  0.43           \\ 
    {\rm V14 } &7066 & 09:34:50.55  & +17:01:45.7 &~{\rm AC}  &0.612 & ~~443.945 & 22.53 & 22.00 & 0.69 &0.43 \\
\hline
\end{tabular}\\
Notes:\\
 (a) last 4 digits of the SDSS identification number, the complete number is obtained by adding: 587742061055830000\\
 (b) RA, Dec coordinates from the SDSS catalogue\\
 (c) RA, Dec coordinates read from the 11084\_17 WFPC2 LEO~T image in the HLA-STSCI\\
\end{table*}

The identification and the main characteristics of the Leo~T variables
are summarized in Table~\ref{tab2}.  The variable stars are ordered by increasing distance from the center of the galaxy published by
Irwin et al. (2007).  Columns 3 and 4 provide the right ascension and
declination (J2000 epoch), respectively. The coordinates were obtained
by cross-matching our sources with the SDSS catalogue.  The SDSS IDs
of the variables stars are provided in Column 2.  Columns 5, 6, and 7
give the variability type (see Section 3.2, for the classification of
the brighter variables), the period, and the Heliocentric Julian date
of maximum light. Columns 8 and 9, give the intensity-weighted mean
$V$ and $I$ magnitudes, while Columns 10 and 11 list the amplitude in
the $V$ and $I$ bands.  Uncertain values for the mean magnitudes and
amplitudes, due to the incompleteness of the light curves, are flagged
by a colon.
\begin{table}
\begin{center}
\caption{Johnson-Cousins $V,I$ photometry of the Leo~T variable stars}
\label{tab3}
\vspace{0.5 cm} 
\begin{tabular}{ccccc}
\hline
\hline
\multicolumn{5}{c}{Leo~T - Star V1 - {\rm AC}} \\ 
\hline
{\rm HJD}                 & {\rm V}           & & {\rm HJD }                & {\rm I}\\ 
{\rm ($-$2454000)} & {\rm (mag)}  & & {\rm ($-$2454000)} & {\rm (mag)}  \\
\hline
 395.171606 &  23.02  &&    403.364745 &  22.10      \\
 395.187579 &  22.97  &&    403.430029 &  22.13      \\
 395.237584 &  22.94  &&    403.446003 &  22.20      \\  
 395.253558 &  22.98  &&    403.496702 &  22.21      \\  
 395.304257 &  22.92  &&    403.512676 &  22.19      \\   
 395.320231 &  23.03  &&    403.563376 &  22.23      \\   
 395.370930 &  22.95  &&    403.579349 &  22.25      \\   
 395.386903 &  22.96  &&    403.630049 &  22.31      \\   
 440.867040 &  22.65  &&    403.646022 &  22.30      \\ 
 440.883014 &  22.74  &&    442.864446 &  22.30      \\  
\hline  		
\end{tabular}
\end{center}
Table~\ref{tab3} is published in its entirety in the electronic edition of the Journal. 
A portion is shown here
for guidance regarding its form and content.
\end{table}         
The Johnson-Cousins $V,I$ time series data of the variable stars are
provided in Table~\ref{tab3}.  They were obtained by calibrating the
F606W and F814W data through a procedure that properly takes into
account the variation in color of the variable stars during the
pulsation cycle.

A finding chart of the Leo~T variable stars is presented in
Figure~\ref{figmapHST4}.  In the map we have also marked 
the RGB stars measured spectroscopically by Kirby et al. (2008).
The map shows that the spatial distributions
of the variables,  RGB stars and gas in Leo~T are significantly off-centered
with respect to the center of the galaxy defined by Irwin et
al. (2007; red cross in Fig.~\ref{figmapHST4}).  The largest fraction
of galaxy stars and all the variables are confined within the cameras
3 and 4 of the WFPC2.  Of the variable stars, only V14 in the camera 4
of the WFPC2 falls slightly outside the galaxy half-light radius.  The
center of the variable star distribution seems to be in-between Irwin
et al. center for the galaxy, and Ryan-Weber et al. (2008) center of
the HI distribution (blue cross in Fig.~\ref{figmapHST4}). We will
further discuss this point in Section 6.

\subsection{The Leo~T RR Lyrae star: properties and distance}
The period of the RR Lyrae star suggests an Oosterhoff-Intermediate (Oo-Int)
classification\footnote{The Galactic GCs divide into two different
  groups depending on the mean period of their ab-type RR Lyrae stars:
  Oosterhoff type I (Oo I) clusters have $\langle P_{ab} \rangle=0.55$
  d, whereas type II (Oo II) clusters have $\langle P \rangle=0.65$
  d. Field and cluster RR Lyrae stars in external galaxies, instead,
  generally have a $\langle P_{ab} \rangle$ intermediate between the
  two Oosterhoff types, hence, they are classified as Oo-Intermediate
  (Oo-Int).}  for the Leo~T UFD, with caution, though, as the
classification relies on just one single object.  We have studied so
far the variable star population of other 7 MW UFDs (see the Introduction section).  
 Six of them were found to contain RR
Lyrae stars with an average pulsation period resembling that observed
in the Oosterhoff type II (Oo~II) MW GCs (Oosterhoff 1939).  The only
exception was, so far, CVn I, the brightest of the MW UFDs and also
the most similar to the classical dSphs, which instead appears to be
Oo-Int (Kuehn et al. 2008), as also are, in fact, the vast majority of
the classical dSphs (see e.g. Clementini 2010).  Leo~T would thus
become the second UFD showing Oo--Int properties, after CVn~I.

We have derived an estimate of the distance to Leo~T from the mean
magnitude of the galaxy RR Lyrae variable (star V6). The mean
magnitude of V6 is ${\rm \langle V\rangle}$ = 23.59 $\pm 0.05$ mag
(where the error takes into account both the internal error of the
photometry and the uncertainty of the photometric calibration).  We
adopt a linear luminosity-metallicity relation for the RR Lyrae stars,
with the slope value estimated by Clementini et al. (2003a) and
Gratton et al. (2004), and the zero point value consistent with a
distance modulus for the LMC of 18.52$\pm$0.09 mag (Clementini et
al. 2003a), i.e.  ${\rm M_{V}(RR)=(0.214 \pm 0.047)\,[Fe/H]+(0.86  
\pm  0.09)}$\footnote{The zero point of the RR Lyrae luminosity--metallicity 
relation is still an open issue. We have adopted here the zero point of 
Clementini et al. 2003a, to be consistent with our latest papers on  the
variable stars in the UFDs, but different choices are possible, see,  
e.g., Cacciari \& Clementini 2003; Benedict et al. 2011.} 
Using for the RR Lyrae star the mean metal abundance of
the galaxy RGB stars measured spectroscopically $\langle {\rm [Fe/H]}
\rangle$=$-1.99 \pm 0.05$, $\sigma_{\rm [Fe/H]}$=0.52 dex (Kirby et
al.  2008, 2011), and a reddening value of $E(\bv)$=0.031$\pm$0.026
mag, from the Schlegel et al. (1998) maps, we find a distance modulus
of 23.06 $\pm$ 0.15 mag (corresponding to a distance d=409
$^{+29}_{-27}$ kpc), in very good agreement with the distance modulus
of 23.1 mag adopted for the galaxy by Irwin et al. (2007).  Here, the
errors include uncertainties in the photometry, reddening,
metallicity, and in the slope and zero point of the RR Lyrae absolute
magnitude versus metallicity relation.
In the previous discussion we have not considered any
evolution off the ZAHB of the RR Lyrae star. At such low metal
abundance,  evolutionary effects could make an RR Lyrae star up to 0.1
mag brighter than the ZAHB luminosity level (see, e.g., Caputo 1997), thus making in turn the
distance modulus as long as $\sim$ 23.2 mag.  

\subsection{Variable stars brighter than the HB: classification, metallicity and distance}
The 13 variables brighter than the RR Lyrae star are located in a
region of the CMD roughly corresponding to the classical instability
strip of bright pulsating stars, however, their periods and
luminosities do not allow to distinguish whether they 
are short period Classical Cepheids (SPCs) or ACs. Furthermore, among them V3 and V7 
have very small and nearly  
equal amplitudes of the $V$ and
$I$ light curves, a characteristic neither observed nor
expected in pulsating stars, which makes V3 and V7 to distinguish significantly from the 
other  variable stars we have detected in Leo~T. We suspect that they are not pulsating variables but binaries.
Unfortunately, with the present  data we cannot reach a firm conclusion on the actual nature of V3 and V7, 
and will thus no further consider them 
in the following discussion on the classification of the Leo~T bright 
variables.

In order to properly classify  the 
 bright variables,  
 in the left panel of Fig.~\ref{figCMD2}  we have compared their position in the
CMD with the Zero Age Horizontal Branch (ZAHB) of evolutionary models for metal abundances $Z$=0.0002 ([Fe/H]=$-$2.0 dex 
\footnote{We have adopted for the conversion from $Z$ to [Fe/H] the
  simple relation: [Fe/H]= $\log{Z/Z_{\odot}}$, where
  $Z_{\odot}$=0.02.};  solid line), 0.0004 ([Fe/H]=$-$1.7 dex;
red-dashed line), and 0.0006 ([Fe/H]=$-$1.5 dex; green dot-dashed line),
respectively, and masses in the range of 0.65 to 2.0 M$_{\odot}$ taken
from Cariulo et al. (2004).  These ZAHBs have been overplotted on the
CMD assuming $E(B-V)$=0.03 mag and a distance modulus of 23.1 mag (as
adopted by Irwin et al., 2007, and in agreement with the value
inferred from the RR Lyrae star).  In addition, in the right panel of
Fig.~\ref{figCMD2} we also show the He burning evolutionary tracks
predicted for Z=0.0002 and masses ranging from 1.6 to 3.0 $M_\odot$.
   Inspection of this plot shows that the bright pulsating variables
  are well reproduced by 1.6--2.0 $M_\odot$ He burning models
  originating from the ZAHB turnover (see below), and therefore having
  ignited the He burning in a partially degenerate core. On the other
  hand, the blue loops of higher stellar masses (2.5 and 3.0 $M_\odot$
  in the figure) are brighter than the variables observed in Leo~T.
  These comparisons, and the SFH presented in Section 5, seem to
  confirm that these variables are more likely massive horizontal
  branch pulsating stars, hence ACs, rather than short period
  Classical Cepheids (SPCs) on the blue loops of a younger (t $\sim$ 50--200 Myr) 
  population. Indeed, it has been suggested by
  several authors (Dolphin et al. 2002, 2003; Clementini et al. 2003b;
  Cordier et al. 2003; Marconi et al. 2004; Caputo et al. 2004;
  Fiorentino et al. 2006) that the ACs are the natural extension of
  the Population I Classical Cepheids to lower metal contents and
  smaller masses, but see also Fiorentino \& Monelli (2012) for a 
  recent discussion of this topic.
  From the evolutionary point of view, there is a
  general consensus that the ACs are central He-burning stars with a
  mass around 1.5 $M_{\odot }$. Indeed, stellar evolution theory (see
  e.g. Castellani \& Degl'Innocenti 1995; Caputo 1998; Fiorentino et
  al. 2006) has shown that for metal abundances $Z \le$ 0.0004 and for
  not-too-old ages ($\le$2 Gyr) the effective temperature in ZAHB
  models, which normally decreases with increasing the mass, reaches a
  minimum at $\log ~ T_{\rm e}\sim 3.74$ (for $Z$=0.0001) or $\sim
  $3.72 (for $Z$=0.0004) around 1-1.2 $M_{\odot }$. By further
  increasing the mass above this value, both the luminosity and the
  effective temperature of the ZAHB structure increase, producing a
  ``ZAHB turn-over'' and an ``upper horizontal branch'' which
  intersects the instability strip again at a luminosity higher than
  the RR Lyrae level (see Fiorentino et al. 2006, and references
  therein). The effect of the higher luminosities on the periods is
  somehow ``balanced'' by the larger masses, and consequently these
  bright massive pulsators show periods that are not significantly
  longer than those typical of RR Lyrae stars, in agreement with the
  observed behavior for the candidate ACs in Leo T.  The comparison in
  Fig.~\ref{figCMD2} also shows that the Leo~T ACs are best fitted by
  the $Z$=0.0002 ZAHB, that very well sets the lower envelope of the
  distribution, while the $Z$=0.0006 ZAHB is too bright, and does not
  extend in color enough to reproduce the observed distribution.  We
  also note that the difference in magnitude between the faintest
  massive pulsators (ACs or SPCs) and the RR Lyrae stars is predicted
  to be a function of metallicity 
(see Fig. 7 in Caputo et al. 2004).  
Inspection of Fig.~\ref{figCMD2} suggests
  that in the case of Leo T this magnitude difference is nicely
  reproduced for $Z$=0.0002 with the exception of only the faintest of the bright variables, 
 star V12. 
  The position of
  this object cannot be explained in terms of differential reddening
  because the reddening law would shift this object to the left of
  the other ACs and a significant amount of extinction would be necessary
 to bring it at the same luminosity level. On the other
  hand, if the star were more metallic (Z $\simeq$ 0.0006) it would
  not be an AC, whose predicted maximum metallicity is Z=0.0004,  and it 
  would be too blue to belong to the blue loop locus of
  SPCs.  Furthermore, it is unlikely that V12 is a foreground RR Lyrae-type pulsator, since at 
  the distance of $\sim$ 290 kpc implied by the star's luminosity  would neither belong to Leo~T or to our own Galaxy.
  Perhaps, V12 is an AC-like pulsator originated from the merging of two
  old low mass stars.  We recall that the origin of the AC
  pulsators is, in fact,  still a matter of debate in the literature, with the two 
  most widely accepted interpretations being that they either represent
  intermediate-age ($<$5 Gyr ) single stars produced by relatively
  recent star formation, or that they formed from mass transfer in
  binary systems as old as the other stars in the parent galaxy.

As discussed in a number of  theoretical papers (see, e.g., Di Criscienzo,  Marconi, \& Caputo,  2004; Marconi et al., 2004)   
 nonlinear convective pulsation models are able to provide firm predictions of the location of the blue boundary of the instability strip, 
 corresponding to the first-overtone mode blue edge. Therefore, by comparing the theoretical predictions for the metallicity of the Leo~T  candidate 
 ACs with the observed distribution of these pulsators in the $M_V$ vs $\log{P}$ diagram we can derive an independent estimate of the Leo~T  
 distance modulus (see Caputo et al. 2000, for a detailed description of this technique).  
 This procedure has 
 been applied in
 Fig.~ \ref{AC_strip} where the  boundaries of the AC instability strip, namely the first overtone blue edge (FOBE) and the fundamental-mode red 
 edge (FRE)\footnote{This is a more uncertain prediction due to its strong dependence on the adopted treatment of convection.} 
 predicted by the pulsation models of  Marconi et al. (2004) have been compared with the observed distribution of the Leo~T candidate AC stars on the assumption of 
 a reddening value of $E(B-V)=0.03$ mag.   By fitting the position of the bluest candidate AC (star V10) to the FOBE line a
 good match between the theoretical and the observed blue edges is
 found with all the candidate  ACs   (open circles; as anticipated we do not  considered V3 and V7 in this comparison) located well inside the
 theoretical strip,  for a distance modulus of 23.05$\pm$0.10 mag, in excellent
 agreement with the value obtained from the Leo~T RR Lyrae star. Here, the error takes into account both the uncertainty in the reddening and in the 
 $M_V - \log P(FOBE)$ relation.   
 
The ACs detected in Leo T cover a period range from $\sim$ 0.5 to less
than 1 day and show a spread of about 0.3 mag in visual
magnitude. Pulsation models of this kind of variables (Marconi et
al. 2004) suggest that near-infrared observations are needed in order
to use them as standard candles through application of $PL$ relations,
and that accurate colors are needed in order to apply optical $PLC$ or
Wesenheit relations to infer individual distances. Furthermore, these
relations are predicted to be mass dependent at variance with the
behavior of Classical Cepheids (see Marconi et al. 2004).  For these
reasons we believe that further more accurate multi-band observations
of Leo~T are required in order to use these pulsators as independent
distance indicators through application of the above--mentioned
relations. On the other hand, it is comforting that the comparisons of
theory versus observations shown in 
Figs.~\ref{figCMD2} and ~\ref{AC_strip} provide distance moduli
that are fully consistent with each other, and in very good agreement
with the modulus inferred from the Leo~T RR Lyrae star (see Section
3.1).  
Finally, we note that the presence of ACs in Leo ~T further
strengthen the similarity with the CVn~I UFD, which also contains a
number of candidate ACs (Kuehn et al. 2008).
\clearpage

\section{The CMD of Leo~T}
\begin{figure*}[h] 
\centering \includegraphics[width=12cm]{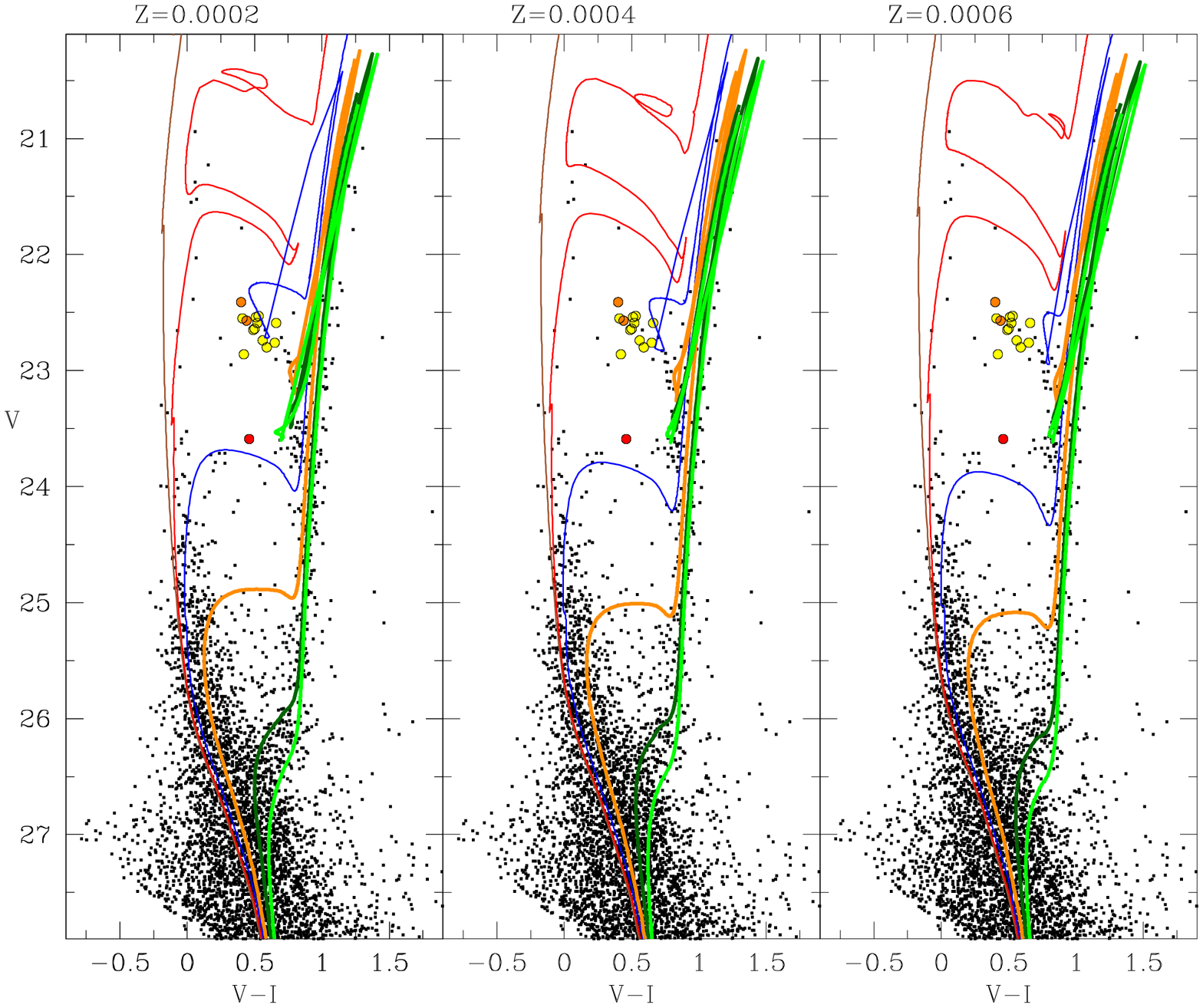}
\caption{Observed CMD with superimposed the Pisa stellar isochrones (Cariulo et al. 2004)
  for metallicities (from left to right) $Z=0.0002$, $Z=0.0004$ and
  $Z=0.0006$ corrected for a distance modulus $(m-M)_0=23.1$ mag and a
  foreground reddening $E(B-V)=0.03$ mag. For each metallicity the
  isochrones are for populations of ages 0.05, 0.2, 1, 3, 8, 12
  Gyr, with the youngest isochrone being the brightest. The RR
  Lyrae star is marked by a (red) filled circle, the bright variables above the HB by (yellow) filled circles, the two candidate binaries by (orange) filled circles.}
\label{isos} 
\end{figure*}

Our CMD for Leo~T reaches $V \sim$ 29 mag, allowing to follow the galaxy's main
sequence (MS) well below the turn-off (TO) of the oldest stars, that
is identified between $V=26.50$ and $V=27.00$ mag (see
Figs.~\ref{figCMD1} and ~\ref{isos}). Several structures are clearly seen: a blue plume
(BP) extending up to $V\sim 21$ mag, a slightly enlarged RGB, a clear
red clump (RC)  between $V\sim 23.5$ and $V\sim 23.8$ mag, a rather 
elongated blue-loop (BL), a well defined
sub-giant branch (SGB), and a putative HB around $V\sim 23.7$ mag
($\sim 23.6$ mag, from the mean magnitude of the RR Lyrae star). These
features reveal the presence of stellar populations of all ages. The
width of the RGB may imply some chemical enrichment, as also suggested
by the spread in metallicity of the RGB stars measured
spectroscopically ($\sigma_{\rm [Fe/H]}$=0.52 dex, Kirby et al. 2008,
2011\footnote{The weighted average of Kirby et al. (2008) individual
  metal abundances for the 10 RGB stars falling in the FOV covered by
  the WFPC2 observations (green filled circles in Fig.~\ref{figCMD1}) is
  $\langle {\rm [Fe/H]} \rangle$ = $-$1.89 with a dispersion of
  $\sigma_{\rm [Fe/H]}$=0.28 dex}).  Signatures of youth are the
bright MS and the BL phase, while the prominent RC is distinctive of
an intermediate-age population, as also suggested by the large number
of ACs detected in Leo~T.  Finally, the mere presence of an RR Lyrae
star and the putative HB indicate stars older than 10 Gyr.
\clearpage
\clearpage
%
 In Figure \ref{isos} we have overplotted on the Leo~T CMD the Pisa
 isochrones (Cariulo et al. 2004) for metallicities $Z=0.0002, 0.0004,
 0.0006$ and ages $0.05, 0.20, 1, 3, 8, 12 $ Gyr. The adopted
 foreground reddening and distance modulus are $E(B-V)$=0.03 mag and
 $(m-M)_0=23.1$ mag, respectively, in agreement with the distance
 moduli derived from the pulsating variable stars. The following conclusions can
 be drawn: 1) the oldest isochrones for the metallicity Z=0.0002 and
 Z=0.0006 bracket well the RGB width; 2) regardless of the
 metallicity, all models seem to (slightly) overestimate the
 luminosity of the RC/HB, suggesting that the adopted distance modulus
 sets a lower limit to the galaxy distance; 3) only a few SGB stars
 are located between isochrones of ages 10 Gyr and 12 Gyr, while most
 of them are consistent with younger ages; 4) above $V\sim 24$ mag, only
 very few BP stars appear bluer than the 200 Myr old isochrones,
 suggesting a low star formation activity in the last 200 Myr. This is
 consistent with the lack of Classical Cepheids in Leo~T. The 200 Myr
 isochrone matches well the clump of stars around $V=21.5$ mag and
 $V-I\approx 0$ mag, indicating that these stars are BL stars. These stars
 were checked carefully, and none of them showed a significant light
 variation over the time-span covered by the present data.

\section{Leo~T SFH}

The determination of the full functional form of the SFH is
accomplished with the population synthesis method described in Cignoni
et al. (2011) and Cignoni \& Tosi (2010).  The underlying stellar
evolution tracks are taken from the Pisa Evolutionary Library (Cariulo
et al. 2004). The core of the method consists of two main steps: 1) a
Monte Carlo procedure, in which a basis set of partial CMDs,
characterized by a step SFH, is generated and convolved with data
uncertainties as derived from extensive artificial star tests. Any
model CMD is written as a linear combination of this basis; 2) a
data-model comparison, in which the level of likelihood between data
and model is assessed in a least-square sense. This point is addressed
minimizing a $\chi^2$, function of the differences in the number of
stars in a set of suitable regions of the CMD. Size and morphology of
the regions are motivated by the photometric errors, Poisson
statistics and stellar population involved. The optimal CMD binning
scheme is found by trial and error, through several combinations of
$\Delta V$ and $\Delta (V-I)$. Given the small size of the sample, a
CMD grid of 0.25 mag in $V$ and 0.1 mag in $V-I$ was found as the most
effective to keep the Poisson fluctuations low and preserve the time
resolution. Once the best SFH is recovered, the statistical
uncertainty is evaluated by a bootstrap technique.

To limit the number of free parameters, all synthetic CMDs were
generated with: 1) a Salpeter initial mass function (IMF); 2) a 30\% binary
fraction \footnote{Primary and secondary masses are randomly extracted
  from the same IMF}; 3) a foreground reddening of $E(B-V)=0.03$ mag,
according to the Schlegel et al.  (1998) maps; 4) a random metallicity
between $Z=0.0002$ and $Z=0.0004$, motivated by the observed magnitude
difference between ACs and the RR-Lyrae stars.  As additional free
parameters, the distance modulus and the differential reddening
$E(B-V)$ were varied in the ranges of 23.00 -- 23.2 mag and 0.0 -- 0.05 mag,
respectively.

\begin{figure*}[t] 
\centering 
\includegraphics[width=12cm]{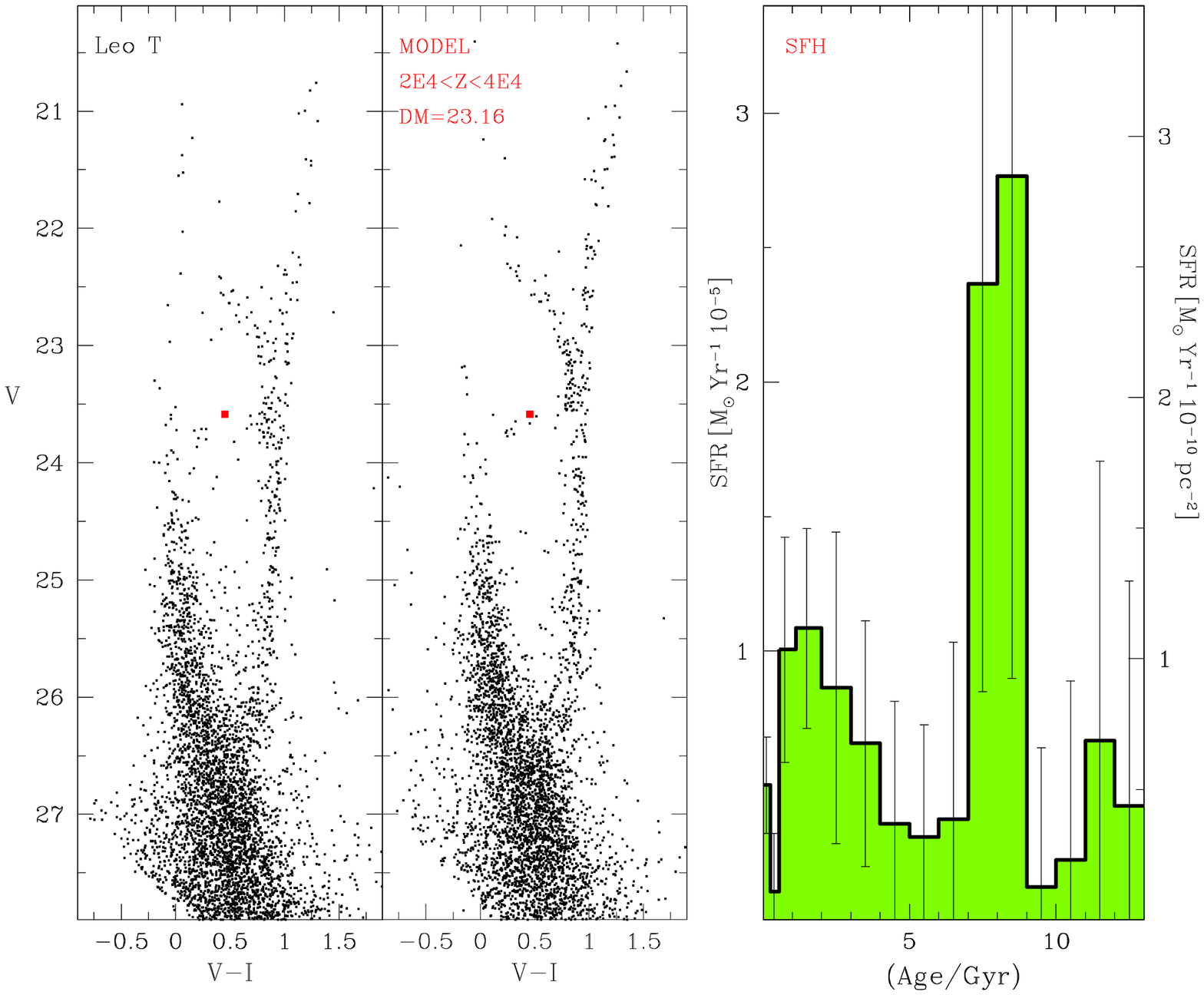}
\caption{Left panel: observational CMD. Middle panel: best synthetic
  CMD. Right panel: best recovered SFH using a distance modulus
  $(m-M)_0=23.16$ mag, and $Z$ values in the range of 0.0002 to 0.0004 (see text for details).}
\label{sfh2} 
\end{figure*}

The right panel of Figure \ref{sfh2} shows the best reconstructed
SFH. The large error bars are partly due to the sample size and partly
due to the photometric errors.
The general trend can be summarized as follows:
\begin{itemize}
\item Leo~T has produced stars continuously since the earliest epochs
  with a gasping regime;
\item Two broad peaks of star formation are recognizable around 1.5
  Gyr and 9 Gyr ago; approximately 36\% of the stars were formed in
  the last 6 Gyr;
\item Three significant gaps are visible in the age range of  0.25 to 0.5
  Gyr, 4 Gyr to 6 Gyr, and 9 Gyr to 13 Gyr ago.
\item A low intensity SF activity started less than 250 Myr ago.
\end{itemize}
The best-fit distance modulus is $(m-M)_0=23.16$ mag, while the
differential reddening is $\Delta\,(B-V)=0.02$ mag.

 Figure \ref{sfh2} shows also the corresponding best synthetic CMD (middle
 panel) together with the observational CMD (left panel). MS counts
 and shape, SGB luminosity spread and RGB color spread are found to be
 in excellent quantitative agreement. A difference is seen in the
 location of the synthetic RC, which is somewhat brighter than
 observed, suggesting that the average metallicity of these stars may
 be slightly higher than in the model. However, increasing the
 metallicity will ultimately reduce the magnitude difference between the 
 ACs and the RR Lyrae  star and make it lower than observed. Another issue
 is the paucity of synthetic BP stars above $V\approx 23$ mag, a CMD
 region populated by massive and intermediate-mass stars on the
 main-sequence and, in the most metal poor case, also by brighter BL
 stars. A contamination of foreground stars is ruled out by the lack
 of blue stars above $g=24$ mag in the control field of de Jong et
 al. (2008) Figure 1;  therefore this discrepancy indicates that
   the SF in the most recent bin (0-250 Myr ago) may be taking place
   on timescales shorter than the bin duration (which is imposed by 
   photometric errors and small number statistics).
\begin{figure*}[t] 
\centering 
\includegraphics[width=12cm]{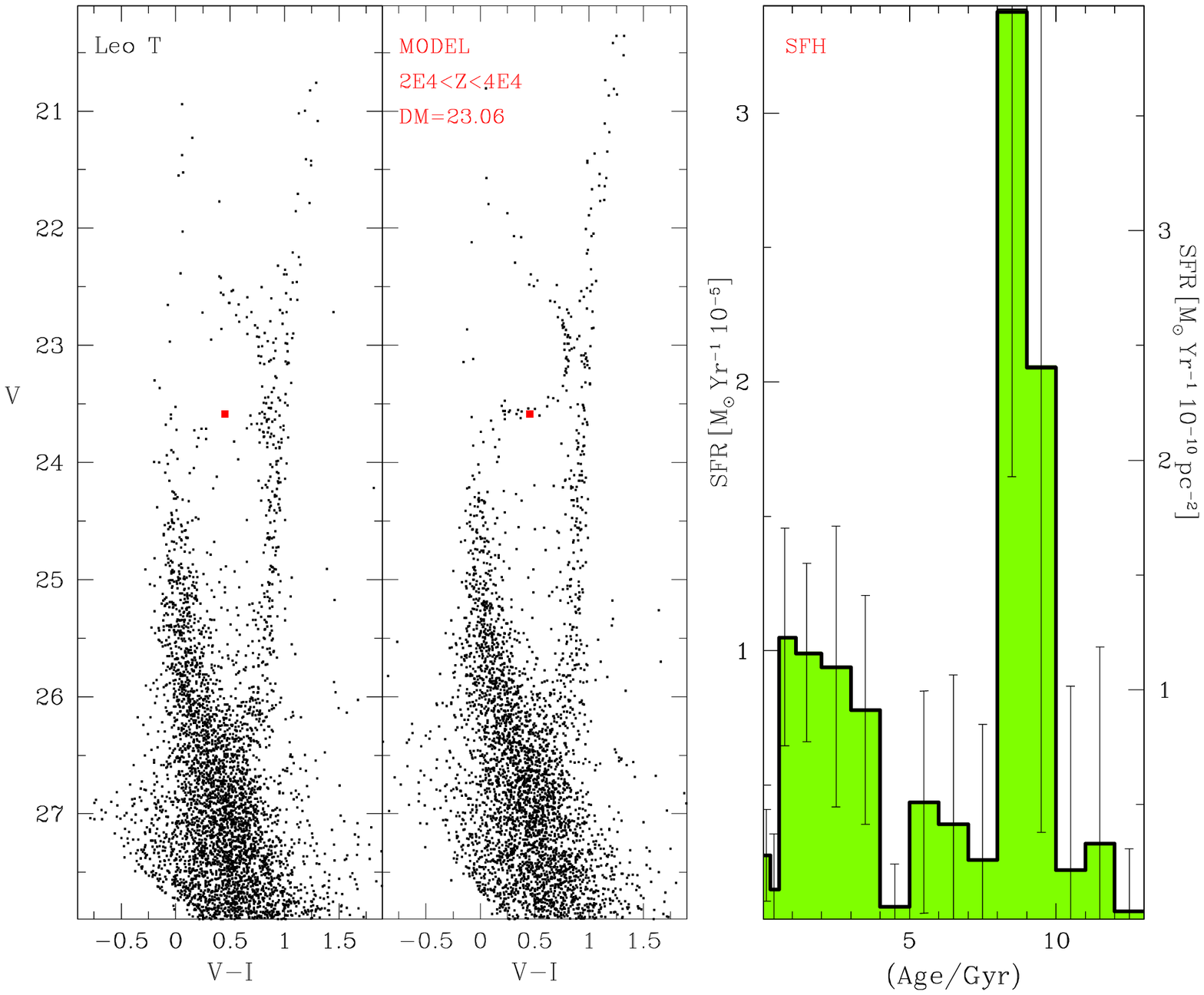}
\caption{Left panel: observational CMD. Middle panel: best synthetic
  CMD. Right panel: best recovered SFH using a distance modulus
  $(m-M)_0=23.06$ mag and $Z$ values in the range of 0.0002 to 0.0004.}
\label{sfh} 
\end{figure*}

To test the robustness of our finding we have re-derived the SFH using
the shorter distance modulus $(m-M)_0=23.06$ inferred from the RR
Lyrae analysis. Figure \ref{sfh} shows the corresponding SFH (right
panel), the best synthetic CMD (middle panel) and the observational
CMD (left panel).  This SFH is fully consistent with the first
one. The only clear difference is the earliest onset of the star
formation activity, that starts earlier for the slightly longer 
distance modulus preferred by the SFH recovery. The match between observational and synthetic CMDs is still
acceptable, although there are differences in the predicted position
of helium burning stars (HB, ZAHB turnover and RC), which are brighter
than the corresponding observational counterparts (RR Lyrae star, ACs, 
and RC stars). In conclusion, the distance modulus which provides the
best fitting SFH is slightly larger but fully consistent with that
resulting from the RR Lyrae and the galaxy's ACs.

\subsection{Comparison with previous SFH studies for Leo T}

de Jong et al. (2008) presented the first study of the Leo~T SFH,
based on their Large Binocular Telescope observations of the galaxy.  More recently, Weisz
et al. (2012) have presented a SFH of Leo~T based on the same dataset
used in this paper.  Our results confirm those of de Jong et
al. (2008) and Weisz et al. (2012), who used similar procedures to
recover the SFH from their CMDs. Like us, those authors found a large
amount of star formation at the earliest ages, followed by a relative
decrease at intermediate ages and then a renewed activity in the last
few Gyr which decreases towards the present time. Interestingly, our
study also revealed a previously overlooked feature, an initial
activity between 10 and 13 Gyr ago significantly lower than at later
epochs. Such a result is mostly driven by the TO population and the
lower envelope of the SGB, CMD features that were simply not
accessible to the shallower LBT sample of de Jong et al. (2008).  On
the other hand, these features are clearly visible in Weisz et al. 's
(2012) CMD, who used the same WFPC2 observations used here. Hence the
discrepancy with Weisz et al. 's (2012) SFH may be due to differences
in the technique used to estimate the SFH. For instance, Weisz et
al. 's (2012) distance, $(m-M)_0=23.05$ mag, and metallicity
($Z$=0.0003-$Z$=0.0005) are derived simultaneously with the SFH, while in
our procedure they are assumed on the basis of the variable star
analysis: although our and Weisz et al. 's (2012) values are
ultimately rather similar, our ``a priori'' assumption on the distance
may have simplified the search for the best SFH.
 
It is also worth of mention that: 1) Weisz et al.'s preferred
   distance of $(m-M)_0=23.05$ mag,  is well consistent with our derived value of
   $(m-M)_0=23.16$ mag, since the former was obtained for an inferred extinction of
   $A_{V}=0.20$ mag, hence, almost a factor two larger than ours; 2) Weizs et
   al's SFH and ours are still consistent within 1-sigma
   errors. Indeed, 50\% of Leo~T's total mass was produced prior to
   7.6 Gyr ago in Weisz et al.'s solution, in good agreement with the 59\%
   produced by our solution prior to 7 Gyr ago. Clearly, in the
 future it would be of great interest to explore the Leo~T UFD with
 deeper, higher-resolution follow-up observations (HST/ACS and
 HST/WFC3).
 
\clearpage

\subsection{Comparison with other dwarfs galaxies}

Overall, and in contrast with the common view that UFDs are pure old
systems, we find that Leo~T experienced a continuous star formation
over most of the past 13 Gyr. In this respect the Leo~T's SFH is a
small-scaled version of the SFH occurring in brighter systems, like the dwarf spheroidals
Carina and Leo~I (gas free), or the dwarf irregular IC~1613 (gas
rich), despite Leo~T being roughly 10 to 100 times less massive (in terms of
luminous mass). Indeed, although the Leo T's stellar production is
slightly peaked prior to 5 Gyr ago, Leo I and IC~1613 are both
characterized by an extended event from about 5 Gyr ago until 2 Gyr ago
(see e.g. Skillman et al. 2003), a shape that is, on average,
comparable with the Leo~T's history. Such a comparison becomes even more
intriguing when we consider Carina: this galaxy experienced a minor
activity prior to 12 Gyr, a dominant burst about 7 Gyr ago (that
lasted about 2 Gyr) and a young burst between 2.5 and 3.5 Gyr ago (see
Hurley-Keller et al. 1998). It is tantalizing to associate these
episodes with the two burst history of Leo~T, keeping in mind that
Leo~T's bursts are spread over a longer interval of time. Moreover, 
while there is evidence for a moderate  chemical enrichment in Leo~T, 
Carina's bursts have different mean metallicities.

A comparison with Transition Type Dwarfs, systems which have cold
gas like Leo~T, but with no or very little star formation, deserves a
separate discussion. To date, only a couple of transition dwarf
galaxies, namely Phoenix and LGS3, have been studied down to the
turn-off of the oldest populations. Although Phoenix and LGS3 are more
massive (according to Hidalgo et al. 2009, 2011 the
total mass in stars is about $4.4\times\,10^{6}\,M_{\odot}$ and
$2.0\times\,10^{6}\,M_{\odot}$, respectively) than Leo~T, all three
galaxies are isolated members of the LG, so tidal interactions are not
likely to have affected their evolution (which is probably driven by
internal physical processes). Phoenix is at roughly the same distance
of Leo~T. The SFH recovered by Hidalgo et al.  (2009) suggests that
both galaxies managed to sustain the star formation for almost a
Hubble time. However two major differences are noteworthy. First, the
fractional mass produced by Leo~T in the last 6 Gyr (36\%) is higher
than in Phoenix (15\%). While feedback from supernovae may be
sufficient to expel gas from their shallow potential wells, the higher
mass of Phoenix may have contributed to keep high its gas density,
allowing to form stars more quickly and thus depleting the gas
reservoirs. Nevertheless, if we accept this view, we are forced to
conclude that dwarf galaxies like Leo~T have probably the lowest
galactic masses able to sustain the star formation for a Hubble
time. Second, Leo~T reached the maximum activity four Gyr later than
Phoenix. Also in this case the lower mass of Leo~T may have played a
role, providing a minor shield from the UV-radiation at the epoch of
the re-ionization, thus delaying the star formation onset.

Located at $0.65\pm\,0.05$ Mpc from us, LGS3 has
a SFH (Hidalgo et al. 2011) showing similarities with both Phoenix and
Leo~T. Like the former it produced a larger amount of mass in the
first 6-7 Gyr, followed by a low level activity which proceeded up to
now. Like the latter, it experienced delayed star formation relative
to Phoenix: LGS3 produced the first 10\% of stars 0.5 Gyr later than
Phoenix, whereas Leo~T produced the same percentage 2-3 Gyr later than
Phoenix.

If Phoenix and LGS3 are dwarfs ``in transition'' between late and
early type galaxies, Leo T may be in a different stage of this
transition. This conclusion is broadly consistent with the SFH
recovered by Weisz et al.  (2011) for another transition dwarf galaxy,
Antlia, which is even more isolated (1.3 Mpc from us) than Phoenix and
LGS3. Although Weisz et al.'s CMD for Antlia does not reach the oldest MS
TO, their SFH is consistent with a constant activity over the entire
Hubble time.

\section{Spatial distribution of stars and gas in Leo~T}

\begin{figure*}[t]
\centering
\includegraphics[width=11cm,clip]{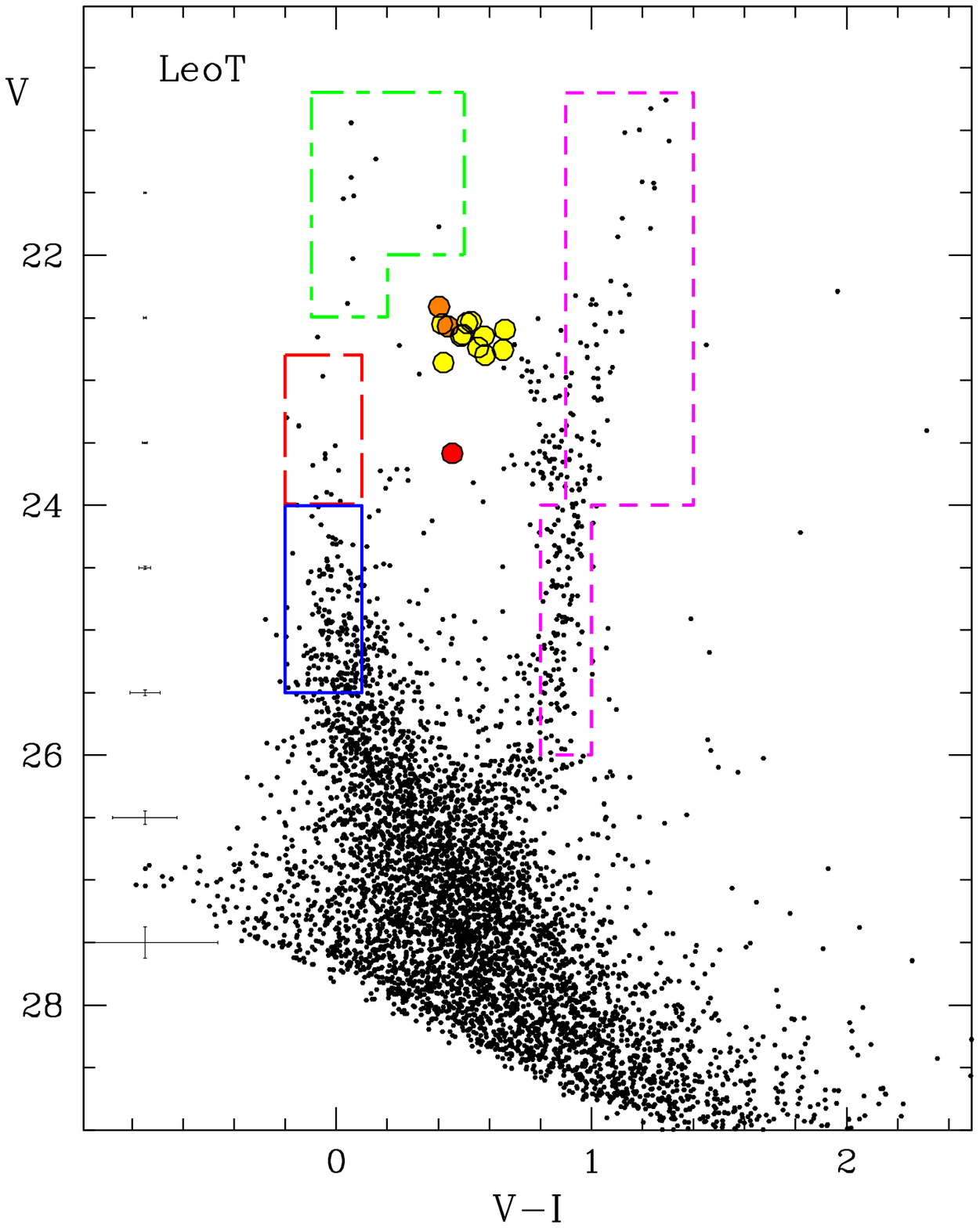} 
\caption{CMD selection of upper-MS (long-dashed, red, box), BL (dot-dashed, green, box), lower-MS 
(solid, blue, box), and RGB stars (dashed, violet, boxes). The bona-fide RR Lyrae star 
and the bright variables above the HB are marked by (red) and (yellow) circles, respectively,  the suspected binaries by (orange) filled circles.}
\label{cmdsel}
\end{figure*}


\begin{figure*}[t]
\centering
\includegraphics[width=11cm,clip]{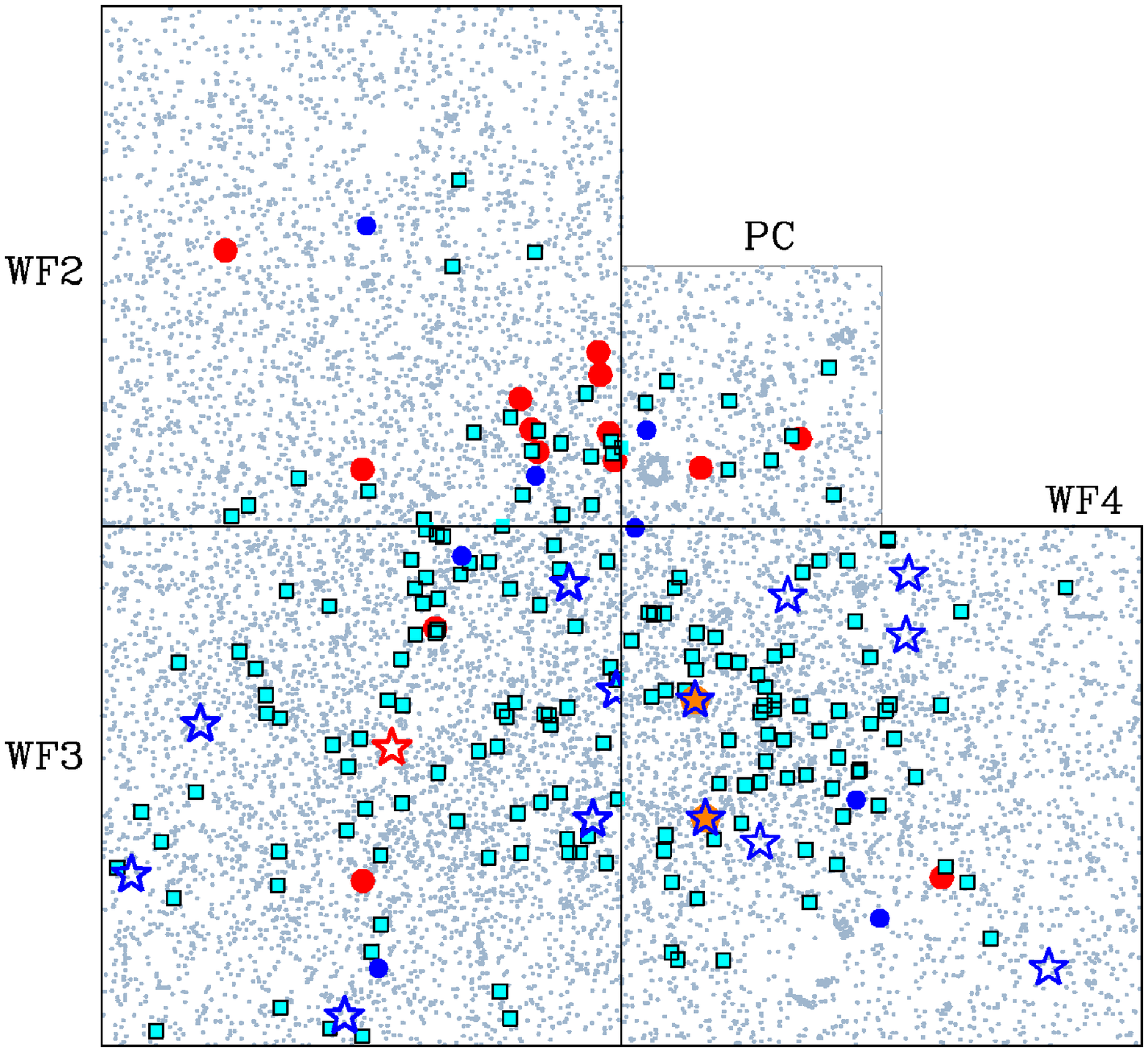} 
\caption{Spatial distribution of the Leo~T stars on the $2.6^{\prime} \times 2.6^{\prime}$ FOV covered by the 4 cameras of
  the WFPC2. (Red) filled circles mark upper-MS stars with 22.8$< V
  \leq$ 24 mag, and $V-I <$ 0.1 mag (long-dashed, red, box selection in
  Fig.~\ref{cmdsel} ); (cyan) squares are lower-MS stars with 24 $< V
  \leq$ 25.5 mag, and $V-I <$ 0.1 mag (solid, blue, box in Fig.~\ref{cmdsel}
  ); (blue) filled circles are BL stars with $V < $ 22 mag, and $V-I <$
  0.5 mag and $ V \leq$ 22.5 mag, and $V-I <$ 0.2 mag (dot-dashed, green, box in
  Fig.~\ref{cmdsel}); (grey) dots show all the remaining stars.  A (red) open 
  star marks the Leo~T RR Lyrae star, the (blue) open stars are the
  bright variables above the HB, the (orange) filled stars are the
  suspected binaries.}
\label{figmapHST42}
\label{pippo}
\end{figure*}

\begin{figure*}[t]\centering
\includegraphics[width=11cm,clip]{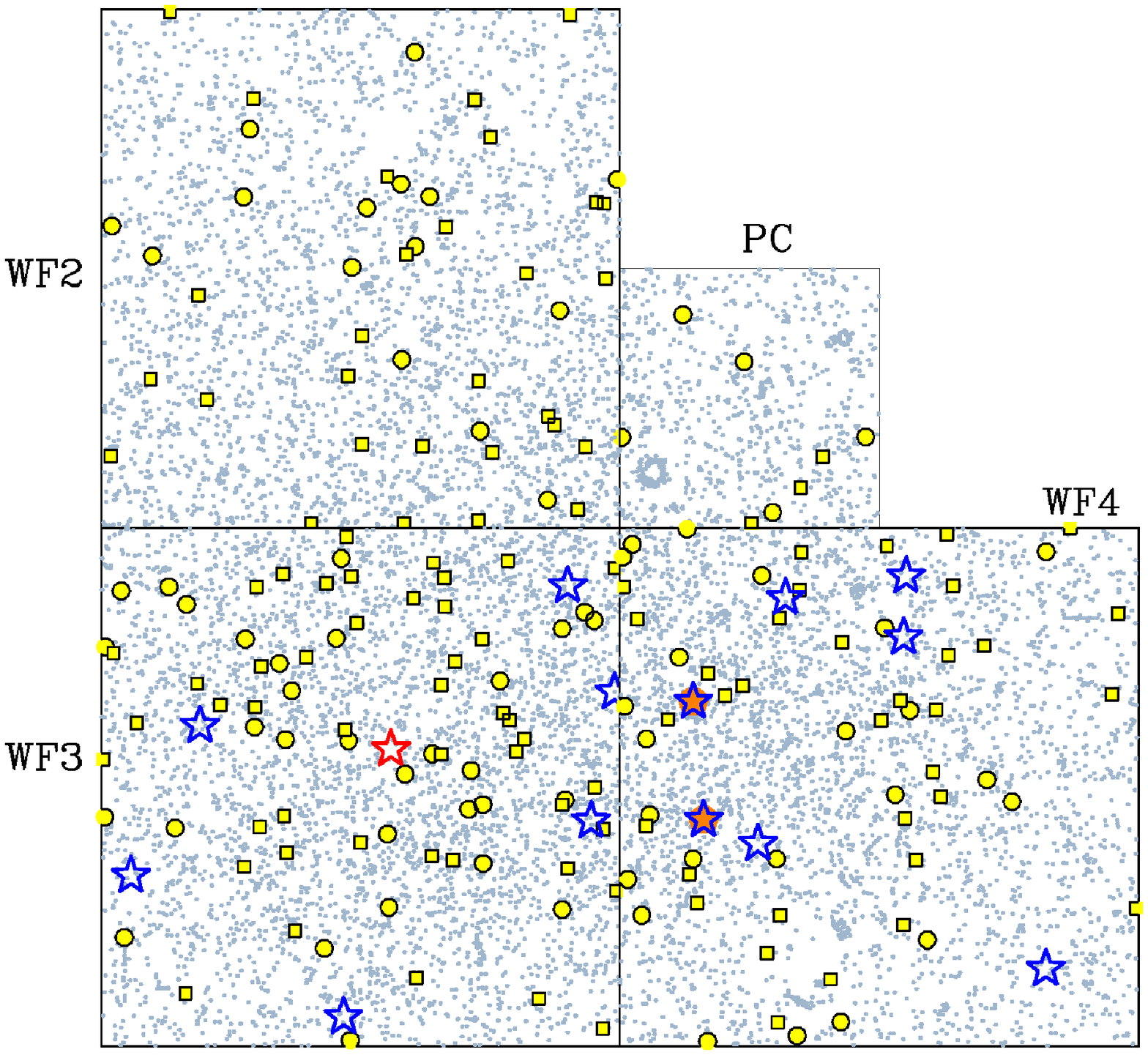} 
\caption{Spatial distribution of the Leo~T stars on the $2.6^{\prime} \times 2.6^{\prime}$ FOV covered by the 4 cameras of
  the WFPC2. (Yellow) filled circles mark RGB stars with $V \leq$ 24
  mag, and $V-I \geq$ 0.9 mag (dashed, violet, upper box in Fig.~\ref{cmdsel});
  (yellow) filled squares mark RGB stars with $24 < V \leq$ 26 mag, and
  $0.8 \leq V-I \leq$ 1 mag (dashed, violet, lower box in Fig.~\ref{cmdsel});
  (grey) dots show all the remaining stars. A (red) open star marks the
  Leo~T RR Lyrae star, the (blue) open stars are the bright variables
  above the HB, the (orange) filled stars are the suspected binaries.}
\label{figmapHST43}
\end{figure*}

The spatial distribution of different stellar populations in a galaxy
provides a direct indication of how and where the star formation took
place at different times. Figure \ref{cmdsel} shows a selection of
upper-MS, BL, lower-MS and RGB stars (above and below the RC), whose
spatial distributions are plotted in Figs.~\ref{figmapHST42} and
\ref{figmapHST43}.
\begin{figure*}[t]
\centering
\includegraphics[width=11cm,clip]{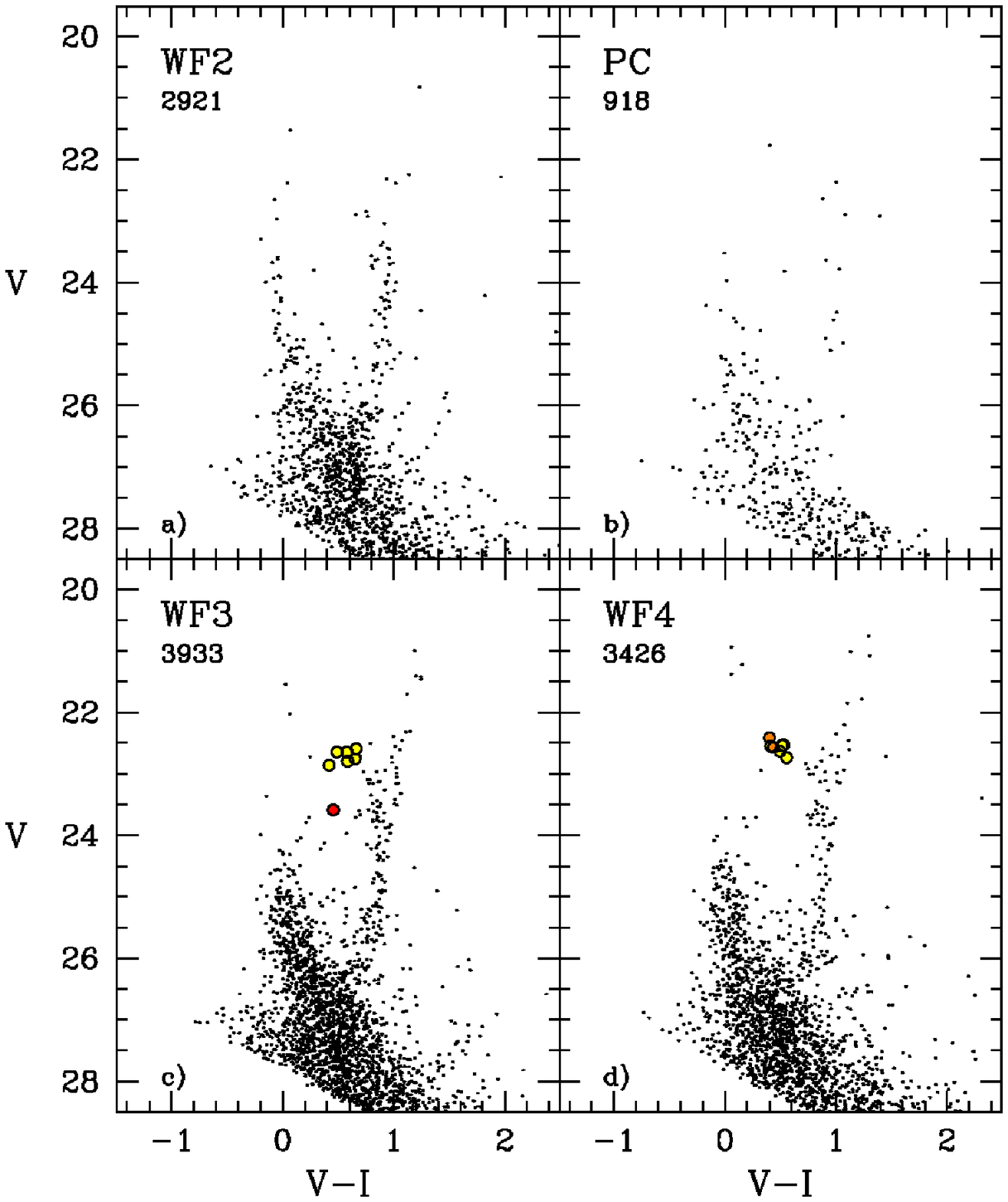} 
\caption{CMDs of Leo~T showing (clockwise from upper left) the stellar
  content of the 4 WFPC2 cameras: WF2, PC, WF4 and WF3,
  respectively. The RR Lyrae star and the bright variables above the
  HB are marked by (red) and (yellow) filled circles, respectively,
  the suspected binaries by (orange) filled circles.}
\label{figcmd4_bis}
\end{figure*}
It can be seen that the distribution of the upper-MS sample (filled, 
red, circles in Fig.~\ref{figmapHST42}) is similar to the BL sample
(filled, blue, circles in Fig.~\ref{figmapHST42}), with both being mostly
concentrated in the WF2 camera. On the other hand, the majority of stars
from the lower MS (filled, cyan, squares in Fig.~\ref{figmapHST42}) as
well as all the variable stars (ACs and RR Lyrae, marked with open, red 
and blue, stars, respectively) are found in the WF3 and WF4
cameras. Finally, the distribution of RGB stars is much more evenly
distributed (see filled, yellow, circles and squares in
Fig.~\ref{figmapHST43}) and only slightly in excess in the  WF3 and WF4
cameras.

In order to get a hint of  the numerical significance of these
differences, we show in Figure \ref{figcmd4_bis} the CMDs of stars in each of the 
 four WFPC2 cameras, separately. 
As expected from the previous discussion, the camera predominantly away
from the central regions (WF2) shows excess counts of BP stars above
$V=24.5$ mag relative to the cameras WF3 and WF4, in spite of being globally
the least populated one. This difference is particularly evident between cameras 
WF2 and WF4, each containing a similar number of RGB, HB and BL
stars, but rather different BP counts and shape. Moreover, the WF4's BP is rather
thick and hooked around $V=24.5$ mag, this is the clear signature of a dominant population
1-2 Gyr old, while the WF2's BP is sharp and extended to bright
magnitudes, consistently with a population not older than 0.2 Gyr. The PC shows, instead,  intermediate properties between the WF2
and WF4 cameras. 

These features suggest that the young stars have a different, asymmetric
distribution with respect to the bulk of older stars which trace the
main body of the galaxy.

Interestingly, the HI surface density (see Fig.~\ref{figHImap}) shows
a similar asymmetry, characterized by a \emph{protrusion} from the main
(round) gas cloud. The causes for this may be manifold, ranging from
an ongoing minor merger and recent accretion of external gas,  to fall
down of gas that once was expelled from the galaxy in a previous star-burst phase. 
Given the proximity of the two features (gas
protrusion and excess of youngest stars in the WF2 camera) one could go a step further and suggest a
recent ``off-center'' propagation of the star formation. Enlarging the
sample by including high spatially-resolved observations of the Leo~T's periphery
would be therefore crucial and urgent to obtain a more complete picture of the  galaxy's history.

\section{Summary and Conclusions}
We have studied the variable stars of the Leo~T
UFD and  performed a quantitative analysis of the galaxy's SFH.
We have detected 14 variables in Leo~T, they include 1 fundamental-mode
RR Lyrae star, 10 ACs,  and two putative binaries.
The average period of the RR Lyrae star ($P$=0.6027 d) suggests  an Oosterhoff-Intermediate
classification for Leo~T, similarly only to CVn~I,  among the UFD galaxies.
The magnitude difference between ACs and the RR Lyrae star suggests a
metallicity lower than Z=0.0004, and more  likely around 0.0002 ([Fe/H]=$-2.0$ dex) for the intermediate-age and the oldest stellar 
components in Leo~T. 
Adopting this metal abundance,  a reddening value $E(B-V)$=0.03 mag, and an  absolute visual magnitude 
for the RR Lyrae star of $M_V(RR)$ =0.44 mag (at [Fe/H]=$-$2.0 dex) 
the distance modulus inferred for Leo~T 
 is $(m-M)_0=23.06 \pm 0.15$ mag, in excellent agreement with  the modulus obtained by fitting 
the First Overtone Blue Edge (FOBE) of the ACs'  sample, $(m-M)_0=23.05 \pm 0.10$ mag, and only slightly shorter
than the value provided by the SFH analysis, $(m-M)_0=23.16$ mag.

In spite of the low mass, Leo T underwent a complex SFH, with two
major star forming episodes, about 7-9 Gyr ago and 1-2 Gyr ago,
overimposed on a continuous star formation activity. The average
  star formation rate per pc$^2$ was around $9\times
  10^{-11}\,M_{\odot}\,\mathrm{yr}^{-1}\,\mathrm{pc}^{-2}$ allover the
  galaxy lifetime, with a range of variation between $\approx
  10^{-11}\,M_{\odot}\,\mathrm{yr}^{-1}\,\mathrm{pc}^{-2}$ and
  $\approx 4\times
  10^{-10}\,M_{\odot}\,\mathrm{yr}^{-1}\,\mathrm{pc}^{-2}$. Overall,
these results are in good agreement with previous determinations
(e.g., de Jong et al. 2008, and Weisz et al. 2012), with the exception
of our lower star formation rate earlier than 9 Gyr ago.

The rather isolated location, and the suggestion that the galaxy would be just falling into
the Milky Way for the first time (Rocha et al. 2012),  might explain why Leo~T
managed to retain its gas and kept forming stars until 1-2 Gyr ago,
for which observational evidence is provided by our identification of
a conspicuous population of AC variables, and even later
(a few hundreds Myr ago), as the presence in the CMD of  upper-MS and BL stars, 
and our reconstruction of the galaxy SFH clearly demonstrate.

Leo~T's youngest stars (upper-MS and BL
stars) are spatially confined in the North-East part of the galaxy
(WF2 camera), whereas intermediate-age (ACs) to old populations (RR Lyrae and RGB stars)
are evenly distributed across cameras WF3 and WF4, and almost absent in
the WF2 camera. Similarly, the distribution of neutral hydrogen
shows an asymmetric spatial distribution, with a low density
protrusion extending to the North-East.
The proximity of 
 youngest stellar component and gas protrusion suggests they are coupled, 
 leaving us with the open question of what triggered this common anomaly in such a rather isolated 
UFD. Observations extending over  a much larger field
 of view will be mandatory to  address this question.

 \bigskip
\acknowledgments 
A special thanks goes to Massimo Dall'Ora for turning our attention to
the UFD galaxies,  since the first discoveries in 2006.
We warmly thank P. Montegriffo for the development
and maintenance of the VARFIND and GRATIS softwares.  Financial
support for this research was provided by COFIS ASI-INAF I/016/07/0,
by the agreement ASI-INAF I/009/10/0, and by PRIN INAF 2010, (P.I.:
G. Clementini).

\end{document}